\documentclass[prb,a4paper,showpacs,twocolumn,superscriptaddress,longbibliography]{revtex4-1}

\usepackage{amssymb}
\usepackage{amsmath}
\usepackage{amsfonts}
\usepackage{graphicx}
\usepackage{bm}
\usepackage{color}
\usepackage{multirow}
\usepackage{natbib}
\usepackage{hyperref}

\renewcommand{\Im}{\,\textrm{Im}\,}

 \DeclareMathOperator{\Tr}{Tr}
\DeclareMathOperator{\tr}{tr}

\DeclareMathOperator{\erf}{erf}

\begin{document}

\title{Inelastic electron scattering off a quantum dot  in the cotunneling regime: the signature of mesoscopic Stoner instability}

\author{E. V. Repin}
\affiliation{Moscow Institute of Physics and Technology, 141700 Moscow, Russia}

\author{I. S. Burmistrov}

\affiliation{L.D. Landau Institute for Theoretical Physics RAS,
Kosygina street 2, 119334 Moscow, Russia}
\affiliation{Moscow Institute of Physics and Technology, 141700 Moscow, Russia}

\begin{abstract}
We explore the inelastic electron scattering cross section off a quantum dot close to the Stoner instability. We focus on the regime of strong Coulomb blockade in which the scattering cross section is dominated by the cotunneling processes. For large enough exchange interaction the quantum dot acquires a finite total spin in the ground state. In this, so-called mesoscopic Stoner instability, regime we find that at low enough temperatures the inelastic scattering cross section (including the contribution due to an elastic electron spin-flip) for an electron with a low energy with respect to the chemical potential is different from the case of a magnetic impurity with the same spin. This difference stems from (i) presence of a low-lying many-body states of a quantum dot and (ii) the correlations of the tunneling amplitudes. Our results provide a possible explanation for absence of the dephasing rate saturation at low temperatures in recent experiment [N. Teneh, A. Yu. Kuntsevich, V. M. Pudalov, and M. Reznikov, Phys. Rev. Lett. {\bf 109}, 226403 (2012)] in which existence of local spin droplets in disordered electron liquid has been unraveled.    
\end{abstract}
\date{\today}


\maketitle

%
\section{Introduction}\label{Sec:Intro}

The electron scattering off a magnetic impurity affects crucially properties of electron systems at low temperatures. The simplest model of a magnetic impurity is a random vector of fixed length equal to $S$. Albeit this model ignores the quantum nature of a spin it is enough to produce interesting nontrivial effects, e.g.  suppression of the superconducting transition temperature due to elastic electron spin-flip [\onlinecite{AG}]. Typically this classical approximation is not adequate for the description of magnetic atoms in real systems since their spin is not large, $S\sim 1$. Importantly, the quantum effects in dynamics of a spin makes electron scattering off a magnetic impurity to be inelastic. For example, the Zeeman splitting makes the spin-flip scattering to be energy dependent and suppresses it due to polarization of the spin along magnetic field [\onlinecite{Vavilov2003}]. The other well-known quantum effect is Kondo renormalization of the interaction coupling between an electron spin and spin of an impurity that leads to nonmonotonic temperature dependence of resistivity (for a review see [\onlinecite{AbrikosovBook}]). 
 
The outcome of interaction between electrons and a magnetic impurity can be conveniently formulated in terms of the scattering cross section. For example, the peculiarity of the Kondo problem can be seen in a nonmonotonic behavior of the inelastic scattering cross section with energy at zero temperature [\onlinecite{Zarand2004}]. This nonmonotonicity is translated into a nonmonotonic temperature dependence of the electron dephasing rate due to rare magnetic impurities. The contribution to the dephasing rate due to inelastic scattering off magnetic impurities affects dependence of the weak localization correction on temperature and magnetic field [\onlinecite{Vavilov2003,Micklitz2006,Borda2007,Micklitz2007,Kettemann2007}]. 

In real materials with Coulomb interaction a magnetic impurity with spin $1/2$ can be formed by an electron occupying a localized level [\onlinecite{SchriefferWolff}]. The magnetic impurity with spin $S>1/2$ can be mimicked by a trap with many electrons localized therein. Recently, such electron droplets with spin $S\approx 2$ (per a droplet) have been detected in two-dimensional (2D) electron system in Si-MOSFET by thermodynamic measurements of a sample magnetization [\onlinecite{Kuntsevich1,*Kuntsevich2}]. In the presence of strong exchange interaction in 2D disordered electron system at low temperatures, the spin of an electron droplet can be finite due to phenomenon of the mesoscopic Stoner instability [\onlinecite{KAA},\onlinecite{Narozhny2000}]. The finite spin of an electron droplet yields the Curie-type behavior of the spin susceptibility. The temperature dependence of measured magnetization is consistent  with the Curie law for the spin susceptibility of a single droplet provided their concentration is inversely proportional to temperature   [\onlinecite{Kuntsevich1,*Kuntsevich2}]. 

Motivated by these experiments [\onlinecite{Kuntsevich1,*Kuntsevich2}] we consider the effect of such many-electron puddles with the finite spin on transport properties of 2D electron system. In particular, we estimate contribution to the dephasing time due to inelastic electron scattering off such droplets at low temperatures ($T$). For a sake of simplicity, we model an electron puddle by a quantum dot described by the so-called universal Hamiltonian [\onlinecite{KAA}] with large charging energy ($E_c$) and ferromagnetic exchange interaction ($J>0$). We assume that the quantum dot is weakly tunnel coupled to electrons participating in transport. 

As a quantum dot is concerned we focus on the regime of strong Coulomb blockade, $E_c\gg T$, with an integer number of electrons on the quantum dot. In this regime the leading contribution to the electron scattering off the quantum dot corresponds to the forth order in the tunneling amplitudes. This is similar to the cotunneling regime in a standard analysis of transport through the strongly Coulomb-blockaded quantum dot. We compare two cases of exchange interaction in the quantum dot: Heisenberg interaction and Ising interaction. In the former case the total spin of the quantum dot in the ground state can be estimated as $S\approx {J}/{[2(\delta-J)]}$ where $\delta$ denotes the mean level spacing for single particle levels of the quantum dot [\onlinecite{KAA}]. Near the macroscopic Stoner instability, $\delta-J\ll \delta, J$, the total spin of the quantum dot is large $S\gg 1$. For the Ising exchange the total spin in the ground state is zero for $J<\delta$, i.e. the mesoscopic Stoner instability is absent [\onlinecite{KAA}]. 

In general, the inelastic cross section consists of three terms: elastic spin-flip, inelastic spin-flip, and inelastic non-spin-flip contributions. In this paper, we concentrate on the case of strong exchange interaction: the quantum dot is close to the macroscopic Stoner instability, $\delta-J\ll \delta$, and low temperatures $T\lesssim \delta-J$. We find that for small energy of incoming electron, $\varepsilon \ll \delta$: 
\begin{itemize}
\item[(i)] the elastic spin-flip contribution is the same as for a magnetic impurity with the spin $S\approx {J}/{[2(\delta-J)]} \gg 1$; 
\item[(ii)] at energies $\varepsilon \gtrsim \delta-J$ the inelastic spin-flip and non-spin-flip channels become active; they add the contribution which is $1/S^2 \sim (1-J/\delta)^2$ smaller than one due to elastic spin-flip. 
\end{itemize} 
The presence of Zeeman splitting which is large in comparison with temperature suppresses the elastic spin-flip contribution due to destruction of the mesoscopic Stoner phase [\onlinecite{BGK2012}]. Then we find that the inelastic cross section vanishes for energies $|\varepsilon|\lesssim \delta-J$. At higher energies $\delta-J \lesssim |\varepsilon| \ll \delta$, the inelastic cross section reaches the value which is of the order of elastic spin flip contribution (without magnetic field) for a magnetic impurity with spin $1/2$. In the case of Ising exchange interaction we find that the inelastic cross section at energies $|\varepsilon| \lesssim \delta-J$ is sensitive to the parity of the number of electrons on the quantum dot: for odd number of electrons there is the elastic spin-flip contribution similar to a magnetic impurity with spin $1/2$. Surprisingly, we find that at energies $\delta-J \lesssim |\varepsilon| \ll \delta$ the inelastic cross section becomes almost insensitive to the parity of the number of electrons. 
 
The paper is organized as follows. In Sec. \ref{Sec:Form} we review the formalism and present the general expression for the inelastic cross section at nonzero temperature. Next (Sec. \ref{Sec:CR}) we apply the general formula and derive the expression for the inelastic scattering cross section for the cotunneling regime. As the simplest example we consider the case of a single-level quantum dot and compare our results to ones obtained before (see Sec. \ref{Sec:Form:1LQD}). Next we consider the inelastic scattering cross section for a many-level quantum dot near the Stoner instability
for Heisenberg (Sec.  \ref{Sec:MLQD}) and Ising (Sec. \ref{Sec:MLQD:Ising}) exchange interactions. We conclude the paper with discussion of relation of our results to the experimentally available setups and with the summary of the main results. Some technical details are given in the Appendices.

\section{Formalism}\label{Sec:Form}

We start with the following Hamiltonian
\begin{equation}
	H = H_{QD}  + H_{R} + H_T.
\end{equation}
Here the first term $H_{QD}$ describes electrons in a quantum dot. We consider a metallic quantum dot, i.e. with the large dimensionless conductance, $g_{\rm Th} = E_{\rm Th}/\delta \gg 1$, where $E_{\rm Th}$ denotes the Thouless energy. In this case, the quantum dot is accurately described by the so-called universal Hamiltonian [\onlinecite{KAA},\onlinecite{ABG}]:
\begin{equation}
H_{QD} = \sum_{\alpha,\sigma} \epsilon_{\alpha\sigma} d^\dag_{\alpha\sigma} d_{\alpha\sigma} + E_c(\hat n- N_0)^2 - J \bm{S}^2 .
\label{eq:H:QD}
\end{equation}
Here $d_{\alpha\sigma}$ and $d^\dag_{\alpha\sigma}$ are the annihilation and creation operators for electrons with an energy $\epsilon_{\alpha\sigma} = \epsilon_\alpha+\mu_B g_L B \sigma/2$ on the quantum dot, where $\sigma=\pm 1$ denotes the spin index, $g_L$ and $\mu_B$  stand for the electron g-factor and the Bohr magneton, respectively. The second term in the right hand side of Eq. \eqref{eq:H:QD} accounts for the Coulomb blockade. It involves the particle number operator,
\begin{equation}
\hat n =  \sum_{\sigma} \hat n_\sigma = \sum_{\alpha} \hat n_\alpha = \sum_{\alpha,\sigma}  d^\dag_{\alpha\sigma} d_{\alpha\sigma},
\end{equation}
and the external charge $N_0$. The last term in the right hand side of Eq. \eqref{eq:H:QD} describes the ferromagnetic Heisenberg exchange interaction ($J>0$). It is expressed via the operator of the total spin on the quantum dot,
\begin{equation}
	\bm{S} =  \frac{1}{2}\sum_{\alpha} \bm{s}_\alpha = \frac{1}{2}\sum_{\alpha,\sigma, \sigma^\prime}  d^\dag_{\alpha\sigma} \bm{\sigma}_{\sigma\sigma^\prime} d_{\alpha\sigma}  .
\end{equation}

We do not consider here interaction in the Cooper channel which are responsible for superconducting correlations in quantum dots [\onlinecite{Schechter2004, Ying2006, *Ying2006b, Schmidt2007, *Schmidt2008, *Alhassid2010, *Alhassid2012, *Nesterov2013, *Nesterov2015}]. 

Next the term $H_R$ describes electrons in a reservoir. For a sake of simplicity, we neglect interaction of electrons in the reservoir and write  the Hamiltonian as 
\begin{equation}
	H_{R} = \sum_{k,\sigma} \varepsilon_{k\sigma} a^\dag_{k\sigma} a_{k\sigma}.
	\label{eq:H:R}
\end{equation}
Here $a^\dag_{\alpha\sigma}$ and $a_{\alpha\sigma}$ are the creation and annihilation operators for electrons with an energy $\varepsilon_{k\sigma}=\varepsilon(k) + \mu_B \tilde{g}_L B \sigma/2$ in the reservoir, where $\tilde{g}_L$ denotes the g-factor in the reservoir. We note that all energies are counted from the chemical potential. 

Finally, the term $H_T$ accounts for the coupling between the quantum dot and the reservoir. We choose it in a standard form of the tunneling Hamiltonian:
\begin{equation}
	H_T = \sum_{\alpha,\sigma,k} t_{\alpha k} d^\dag_{\alpha\sigma} a_{k\sigma} + h.c.
\end{equation}
We emphasize that there is no spin-flip of electron during the tunneling event from the quantum dot to the reservoir or vice versa. In what follows we neglect the effect of electrons in the reservoir on dynamics of the total spin of the quantum dot (see Refs. [\onlinecite{Shnirman2015},\onlinecite{Shnirman2016}]).

Following Ref. [\onlinecite{Borda2007}], the T-matrix for scattering of electrons from the state $|\bm{k}\sigma\rangle $ with energy $\varepsilon= \varepsilon_{k,\sigma}$ to the state $|\bm{k^\prime}\sigma^\prime\rangle$ can be written in terms of the Green's functions:
\begin{equation}
	\langle \bm{k^\prime}\sigma^\prime| \mathcal{T} |\bm{k}\sigma\rangle = - \Bigl [ 
	\underline{\mathcal{G}}^{(0)}_{k^\prime \sigma^\prime}(\varepsilon) \Bigr ]^{-1}\underline{\mathcal{G}}^A_{k^\prime \sigma^\prime;k \sigma}(\varepsilon) \Bigl [ \underline{\mathcal{G}}^{(0)}_{k \sigma}(\varepsilon) \Bigr ]^{-1} .
\label{eq:T:1}
\end{equation}
where $\underline{\mathcal{G}}^{(0)}$ and $\underline{\mathcal{G}}$ are the free and  full many-body Green's functions for electrons in the reservoir, respectively. Using the Dyson equation for the advanced Green's function $\underline{\mathcal{G}}^A$, Eq. \eqref{eq:T:1} can be rewritten as follows
\begin{align}
	\langle \bm{k^\prime}\sigma^\prime| \mathcal{T} |\bm{k}\sigma\rangle =& - \delta_{\bm{k^\prime},\bm{k}}\delta_{\sigma^\prime,\sigma}\Bigl [ \underline{\mathcal{G}}^{(0)}_{k \sigma}(\varepsilon) \Bigr ]^{-1} \notag \\
&	- 
	\sum_{\alpha\beta} \bar{t}_{k^\prime \beta} \mathcal{G}^A_{\beta\sigma^\prime;\alpha\sigma}(\varepsilon) t_{\alpha k} .
	\label{eq:2}
\end{align}
Here $\mathcal{G}^A_{\beta\sigma^\prime;\alpha\sigma}(\varepsilon)$ is the exact advanced Green's function for electrons in the quantum dot. The corresponding Matsubara Green's function $\mathcal{G}^A_{\beta\sigma^\prime;\alpha\sigma}(i\varepsilon)$ can be found in the imaginary time as follows (see e.g., Ref. [\onlinecite{Mahan}]):
\begin{equation}
	\mathcal{G}_{\alpha\sigma;\beta\sigma^\prime}(\tau) = - \frac{1}{\mathcal{Z}} \Tr \Bigl [ e^{-\tau H} d^\dag_{\beta\sigma^\prime} e^{-(\beta-\tau) H} d_{\alpha\sigma} \Bigr ] ,
	\label{eq:3}
\end{equation}
where $\tau>0$, $\beta = 1/T$ and $\mathcal{Z} = \Tr e^{-\beta H}$ stands for the grand canonical partition function. The total scattering cross section for an electron in a state $|\bm{k}\sigma\rangle$ is related with the T-matrix as [\onlinecite{Borda2007}]
\begin{equation}
	\sigma_{\rm tot}^\sigma = \frac{2}{v_F} \Im \langle \bm{k}\sigma| \mathcal{T} |\bm{k}\sigma\rangle .
\end{equation}
Here $v_F$ is the velocity of electrons in the reservoir at the Fermi level. In our problem of electron scattering off the quantum dot it is more convenient to study the following quantity
\begin{equation}
	\mathcal{A}_{\rm tot}^\sigma(\varepsilon) = \sum_{k} \delta(\varepsilon-\varepsilon_{k\sigma}) \Im \langle \bm{k}\sigma| \mathcal{T} |\bm{k}\sigma\rangle ,
\end{equation}
which is the scattering cross section averages with the single-particle density of states in the reservoir. 
Using Eq. \eqref{eq:2}, we can express  the quantity $\mathcal{A}_{\rm tot}^\sigma(\varepsilon)$ as 
\begin{equation}
	\mathcal{A}_{\rm tot}^\sigma(\varepsilon) =  \Im 
	\sum_{\alpha\beta} Q^\sigma_{\beta\alpha}(\varepsilon) \mathcal{G}^A_{\alpha\sigma;\beta\sigma}(\varepsilon) .
	\label{eq:A:gen}
\end{equation}
Here we introduce the matrix 
\begin{equation}
 Q_{\alpha\beta}^\sigma(\varepsilon) = 
 \sum_{k} \delta(\varepsilon-\varepsilon_{k\sigma}) t_{\alpha k} \bar{t}_{k \beta} .
\end{equation}
This matrix characterizes the tunnel junction in the following way. Let us define the matrix $\hat g_{\alpha\beta} = (4\pi^2/\delta) \sum_\sigma Q_{\alpha\beta}^\sigma(\varepsilon)$. Then for an electron with the energy $\varepsilon$ the effective number of open tunneling channels $N_{\rm ch}$ and the effective dimensionless (in units $e^2/h$) channel conductance $g_{\rm ch}$ can be written as  
\begin{equation}
N_{\rm ch} = \frac{(\tr \hat g)^2}{\tr \hat g^2}, \qquad g_{\rm ch} =  \frac{\tr \hat g^2}{\tr \hat g} .
\label{eq:Nch}
\end{equation}
We assume that the total conductance of the tunneling junction is small, $g_T=g_{\rm ch} N_{\rm ch} = \tr \hat g \ll 1$. 

We stress that the T-matrix obtained in accordance with Eq. \eqref{eq:A:gen} is averaged over the equilibrium density matrix of the quantum dot and reservoir. In particular, this averaging involves summation over initial states of the quantum dot with the Gibbs weight. Hence, a standard expression for the elastic scattering $\sigma_{\rm el}\propto |\langle \bm{k^\prime}\sigma| \mathcal{T} |\bm{k}\sigma\rangle|^2$, where $\varepsilon_{k\sigma}=\varepsilon_{k^\prime\sigma}$, is inapplicable for our definition of the T-matrix. In what follows, we shall extract the inelastic part of the cross section directly from the final expression for the total cross section (see Sec. \ref{Sec:MLQD}).

\section{The scattering cross section in the cotunneling regime}
\label{Sec:CR}

To the lowest order in $Q_{\alpha\beta}^\sigma(\varepsilon)$ the scattering cross section is determined by the Green's function of electrons on an  isolated quantum dot, i.e. the Green's function corresponding to the Hamiltonian $H_{QD}$. Then, if quantities $Q_{\alpha\beta}^\sigma(\varepsilon)$ are real, the scattering cross section is determined by the tunneling density of states for the isolated quantum dot. In the case of Coulomb valley, this implies exponentially small scattering cross section at low energies $|\varepsilon| < E_c$. 

To calculate the scattering cross section to the forth order in the tunneling amplitudes let us 
introduce the basis of the exact many-body eigenstates $|i\rangle$ for the Hamiltonian \eqref{eq:H:QD} of the isolated quantum dot: $H_{QD} |i\rangle = E_i |i\rangle$. Then computing the Green's function of electrons on the quantum dot to the second order in tunneling (see Appendix 
\ref{App:A}) we find the following result for the total scattering cross section: 
\begin{gather}
	\mathcal{A}_{\rm tot}^\sigma(\varepsilon) =  \pi [1+e^{-\beta\varepsilon}] \sum_{\alpha\beta\gamma\eta} \sum_{i,f,\sigma^\prime}p_i  \int d\varepsilon^\prime\,  \frac{Q^{\sigma}_{\beta\alpha}(\varepsilon)Q^{\sigma^\prime}_{\gamma\eta}(\varepsilon^\prime) }{1+e^{-\beta \varepsilon^\prime}} \notag \\
	 \times
	\langle i | d^\dag_{\gamma \sigma^\prime} \frac{1}{\varepsilon^\prime-E_i + H_{QD}} d_{\alpha\sigma} +
	d_{\alpha\sigma} \frac{1}{\varepsilon+E_i -  H_{QD}} d^\dag_{\gamma \sigma^\prime}| f\rangle 
	\notag \\
	 \times \langle f | d^\dag_{\beta \sigma} \frac{1}{\varepsilon^\prime-E_i + H_{QD}} d_{\eta\sigma^\prime} + 
	d_{\eta\sigma^\prime} \frac{1}{\varepsilon+E_i -  H_{QD}} d^\dag_{\beta \sigma}| i\rangle 
	\notag \\
	 \times \delta(\varepsilon+E_i - E_f-\varepsilon^\prime) .
	\label{eq:4}
\end{gather}
Here $p_i = \exp(-\beta E_i)/Z$, where $Z=\sum_i \exp(-\beta E_i)$, is the Gibbs probability for the initial states of the quantum dot.
We mention that the result \eqref{eq:4} can also be obtained within the generalized Fermi golden rule approach for the T-matrix (see Appendix \ref{App2}). As discussed above, we will be interested in the inelastic scattering only, which means that we will always be considering different initial and final states of the quantum dot, $i\ne f$. 
In what follows we neglect possible dependence of $Q^{\sigma}_{\beta\alpha}$ on spin index $\sigma$.


\subsection{Single-level quantum dot}
\label{Sec:Form:1LQD}

To illustrate the general expression \eqref{eq:4} for the scattering cross section  we consider  
a simple example of a single level quantum dot. In this case there are four many-body states: 
the state without electrons, $|0\rangle$, two states with single electron, $|\uparrow\rangle$ and $|\downarrow\rangle$, and the state with two-electrons with opposite spins, $|\uparrow\downarrow\rangle$. We note that although the universal Hamiltonian \eqref{eq:H:QD} is not justified for a single level quantum dot, the general expression \eqref{eq:4} written in terms of exact many-body eigenstates is correct. Then, we find from Eq. \eqref{eq:4} 
\begin{align}
	\mathcal{A}_{\rm tot}^\sigma(\varepsilon) & = \pi Q^2(\varepsilon) \left [ \frac{p_0+p_\sigma}{(\varepsilon+E_0-E_\sigma)^2} + \frac{p_{\bar{\sigma}}+p_{\uparrow\downarrow}}{(\varepsilon+E_{\bar{\sigma}}-E_{\uparrow\downarrow})^2} \right ] \notag \\
& + \pi Q(\varepsilon) Q(\varepsilon+E_\sigma - E_{\bar{\sigma}}) \frac{1+e^{-\beta\varepsilon}}{1+e^{-\beta(\varepsilon+E_\sigma - E_{\bar{\sigma}})}} \notag \\
	& \times p_{\sigma} \left [ 
	\frac{1}{E_{\bar{\sigma}}-E_0-\varepsilon} + \frac{1}{\varepsilon+E_\sigma-E_{\uparrow\downarrow}}\right ]^2 .
	\label{eq:SLQD:1}
	\end{align}
Here $\sigma  = \uparrow,\downarrow$ and  $\bar{\sigma}  = \downarrow,\uparrow$, respectively. 
The first term in the right hand side of Eq. \eqref{eq:SLQD:1} describes elastic spin-flip of electron with energy $\varepsilon$ and spin projection $\sigma$ after the scattering off the single level quantum dot. The second term corresponds to the scattering with spin-flip. In the absence of magnetic field the two states with single electron have the same energy,  $E_\uparrow = E_\downarrow$, and the result \eqref{eq:SLQD:1} coincide with the result of Ref. [\onlinecite{KonigGefen1,*KonigGefen2}] for the full transmission probability. In the presence of magnetic field spin up and spin down states are not equivalent, $E_\uparrow \neq E_\downarrow$, and the spin-flip scattering becomes inelastic. In the absence of interaction the energy of the state with two electrons is expressed via the energies of the states with one and zero electrons, $E_{\uparrow\downarrow} = E_\uparrow + E_\downarrow - E_0$. Then, the spin-flip term in the scattering cross section \eqref{eq:SLQD:1} vanishes. The elastic contribution becomes independent of temperature. In agreement with Ref. [\onlinecite{KonigGefen1,*KonigGefen2}], the scattering of electrons off the single level quantum dot becomes fully coherent.   
 	
For $E_0, E_{\uparrow\downarrow} \to \infty$ the single-level quantum dot can be singly occupied only, i.e. the quantum dot behaves as the spin $1/2$. In this case spin-flip inelastic part of Eq. \eqref{eq:SLQD:1} reduces to the 
following expression:
\begin{equation}
\mathcal{A}_{\rm inel,sf}^\sigma(\varepsilon) =
\pi \nu^2 J_s^2 \Bigl [ \bar{p}_\downarrow \bigl [1-n_F(\varepsilon+\omega_\sigma)\bigr ] + \bar{p}_\uparrow n_F(\varepsilon+\omega_\sigma) \Bigr ] .
\end{equation}
Here $\omega_\sigma= E_\sigma-E_{\bar{\sigma}}$, $n_F(\varepsilon) = 1/[1+\exp(\beta \varepsilon)]$ stands for the Fermi-Dirac distribution function, and $\bar{p}_\sigma = n_F(\omega_\sigma)$ is the probability of the state with spin projection $\sigma$.  Neglecting dependence of the tunneling amplitudes in $Q$ on the energy, we can write the effective exchange  coupling between the spin of electrons in the reservoir and the spin of electrons on the quantum dot as $J_s = \nu^{-1} Q [1/E_0+1/E_{\uparrow\downarrow}]$, where $\nu$ is the average density of states per spin projection at the Fermi level for electrons in the reservoir. If we assume that there are many such quantum dots (spin $1/2$ impurities) with the concentration $n_{s}$ and define the spin-flip rate for an electron in the reservoir as 
$({2n_{s}}/{\nu}) \mathcal{A}_{\rm inel,sf}^\sigma(\varepsilon)$,
we reproduce the result of Ref. [\onlinecite{Glazman2003}].


\subsection{Many-level quantum dot near Stoner instability} 
\label{Sec:MLQD}

Now we consider the many-level quantum dot described by the universal Hamiltonian \eqref{eq:H:QD}. We remind that the charging energy $E_c$ is large, $E_c \gg T, \varepsilon, \delta, J$, and the external charge $N_0$ has an integer value. Then, the energy of intermediate states in the right hand side of Eq. \eqref{eq:4} is equal to the charging energy, $H_{QD}-E_i = E_c$. Dropping the elastic contribution, i.e. the term with $|i\rangle=|f\rangle$, from Eq. \eqref{eq:4}, and using the commutation relation $[d_{\alpha \sigma}^\dagger, d_{\beta \sigma^\prime}] =\delta_{\alpha \beta}\delta_{\sigma\sigma^\prime}-2 d_{\beta \sigma^\prime} d_{\alpha \sigma}^\dagger$, we rewrite the inelastic contribution to the scattering cross section as   
\begin{align}
	\mathcal{A}_{\rm inel}^\sigma(\varepsilon) & = \frac{4\pi}{E_c^2} \sum_{\alpha,\gamma} \sum_{f\neq  i,\sigma^\prime} Q_{\gamma\alpha}(\varepsilon)Q_{\alpha\gamma}(\varepsilon+E_i - E_f)\notag\\
&\times	  \frac{p_i[1+e^{-\beta\varepsilon}]}{1+e^{-\beta(\varepsilon+E_i-E_f)}} \langle i | d^\dag_{\alpha \sigma^\prime} d_{\alpha\sigma}  | f\rangle 
	\langle f |  d^\dag_{\gamma \sigma} d_{\gamma\sigma^\prime} | i\rangle  \notag \\
	& + \frac{4 \pi}{E_c^2} \sum_{\alpha\neq\gamma} \sum_{f\neq  i,\sigma^\prime} Q_{\alpha\alpha}(\varepsilon)Q_{\gamma\gamma}(\varepsilon+E_i - E_f)\notag \\
&\times	  \frac{p_i[1+e^{-\beta\varepsilon}]}{1+e^{-\beta(\varepsilon+E_i-E_f)}} \langle i | d^\dag_{\gamma \sigma^\prime} d_{\alpha\sigma}  | f\rangle 
	\langle f |  d^\dag_{\alpha \sigma} d_{\gamma\sigma^\prime} | i\rangle 
		\label{eq:r:t3sf}
\end{align}
Here we take into account that the initial and final states of the quantum dot has the same number of electrons.  

Equation \eqref{eq:r:t3sf} constitutes the main result of our paper. We note that it can be applied to computation of the inelastic cross section for an arbitrary Hamiltonian which describes a quantum dot provided this Hamiltonian conserves the total number of electrons $N$ and energies of the many-body exact states with $N$ and $N\pm 1$ are different by large value of charging energy. For the universal Hamiltonian \eqref{eq:H:QD} the matrix elements of single-particle operators in Eq. \eqref{eq:r:t3sf} can be computed exactly by means of the Wei-Norman-Kolokolov method [\onlinecite{WeiNorman},\onlinecite{KOLOKOLOV1986, *KOLOKOLOV1990, *Kolokolov1994, *Kolokolov1995, *Kolokolov1996}] employed for exact evaluation of the spin susceptibility and tunneling density of states recently [\onlinecite{BGK2010},\onlinecite{BGK2012}]. 
Since in this work we are interested at low temperatures, $T\ll \delta$, and in low energies of an incoming electron, $|\varepsilon| \ll \delta$, we can use the straightforward approach with Clebsch-Gordan coefficients used for description of conductance [\onlinecite{Baranger2003},\onlinecite{Alhassid2003}] and shot noise [\onlinecite{Konig2012}] through a quantum dot with Heisenberg exchange at low temperatures.

In general, the tunneling amplitudes $t_{\alpha k}$ are random quantities due to random behavior of electron wave functions on a quantum dot. In what follows, we are interested in the case when energies of an electron before ($\varepsilon$) and after ($\varepsilon^\prime=\varepsilon+E_i-E_f$) scattering are small in comparison with the Fermi energy of electrons in the reservoir. Thus we can neglect the energy dependence in the quantities $Q_{\alpha\gamma}$. For a metallic quantum dot, $g_{\rm Th} \gg 1$, the averaging of the tunneling amplitudes over disorder realizations can be performed independently of the single particle energy levels $\epsilon_\alpha$. Using the following relations [\onlinecite{footnote1}]
\begin{equation}
	\overline{Q_{\alpha\gamma} Q_{\gamma\alpha}} = \begin{cases}
		Q^2, \quad & \alpha\neq \gamma , \\
		(2/\bm{\beta}) Q^2, \quad & \alpha=\gamma ,
	\end{cases}
\end{equation}
and
\begin{equation}
	\overline{Q_{\alpha\alpha} Q_{\gamma\gamma}} = 
	Q^2, \quad \alpha\neq \gamma  ,
\end{equation}
where the parameter $\bm{\beta} = 1$ and $2$ for the orthogonal class AI and  the unitary class A, respectively.  Then after the averaging of Eq. \eqref{eq:r:t3sf} over disorder we obtain
\begin{align}
{\mathcal{A}_{\rm inel}^\sigma(\varepsilon)}   & = \frac{4 \pi Q^2}{E_c^2} \sum_{f\neq  i}    \frac{p_i[1+e^{-\beta\varepsilon}]}{1+e^{-\beta(\varepsilon+E_i-E_f)}}   \Biggl \{
 \bigl |\langle i| S^{-\sigma} |f\rangle\bigr |^2 \notag \\
& 
+ \left (\frac{2}{\bm{\beta}}-1\right ) \sum_{\alpha,\sigma^\prime}    \bigl | \langle i | d^\dag_{\alpha\sigma^\prime} d_{\alpha\sigma}  | f\rangle \bigr |^2 \notag \\
& +
\sum_{\alpha\neq\gamma,\sigma^\prime}\bigl | \langle i | d^\dag_{\gamma \sigma^\prime} d_{\alpha\sigma}  | f\rangle \bigr |^2 \Biggr \} .
	\label{eq:r:t4a}
\end{align}
Here we take into account that the operator $\hat n_{\sigma}$ does not change the many-body state and the states $|i\rangle$ and $|f\rangle$ are different, $\langle i | \hat n_{\sigma} | f\rangle = 0$.
The first line in Eq. \eqref{eq:r:t4a} corresponds to the contribution to the scattering cross section due to rotation of the total spin of the quantum dot as a whole, i.e. the total spin in the initial and final states are the same. The other terms in Eq. \eqref{eq:r:t4a} arise because in the case of the quantum dot the total spin is composed from spins of individual electrons occupying single-particle levels. These additional contributions increase inelastic scattering cross section off the quantum dot in comparison with a magnetic impurity with the same value of the spin. 

Let us consider the case of an electron with large energy, $\varepsilon \gg E_f, E_i,T$. Then the inelastic scattering cross section becomes  
\begin{align}
{\mathcal{A}_{\rm inel}^\sigma} & = \frac{4 \pi Q^2}{E_c^2} \sum_i p_i \Bigl \langle i \Bigl | S(S+1) -S_z^2-\sigma S_z \notag \\  
 +& \left (\frac{2}{\bm{\beta}}-1\right ) \sum_\alpha \Bigl[ \hat n_{\alpha\bar{\sigma}} (1-\hat n_{\alpha \sigma})  + \hat n_{\alpha\sigma}\bigl (1 - \langle i | \hat n_{\alpha \sigma} |i \rangle \bigr )\Bigr ]
\notag \\  
+ & \frac{1}{2}  \sum_{ \gamma\neq\alpha} \hat n_\gamma (2 - \hat n_\alpha)  
\Bigr | i \Bigr \rangle . 
\label{eq:HE:SCS1}
\end{align}
We note that the last term in Eq. \eqref{eq:HE:SCS1} is proportional to the number $K$ of available single-particle levels. Typically, the increase of an electron energy on $\delta$ adds a new final state of the quantum dot which contribute in the sum in Eq. \eqref{eq:HE:SCS1}. At zero temperature it can be estimated as $K \sim \varepsilon/\delta$. Assuming that $K \gg N_0, S$, we obtain that the inelastic scattering cross section is proportional to the electron energy, $\mathcal{A}_{\rm inel}^\sigma  = 4\pi Q^2 N_0 \varepsilon/E_c^2$, for $\delta N_0, \delta J/[2(\delta-J)] \ll \varepsilon \ll E_c$.

\begin{figure}[t]
	\centerline{a)\includegraphics[width=0.2\textwidth]{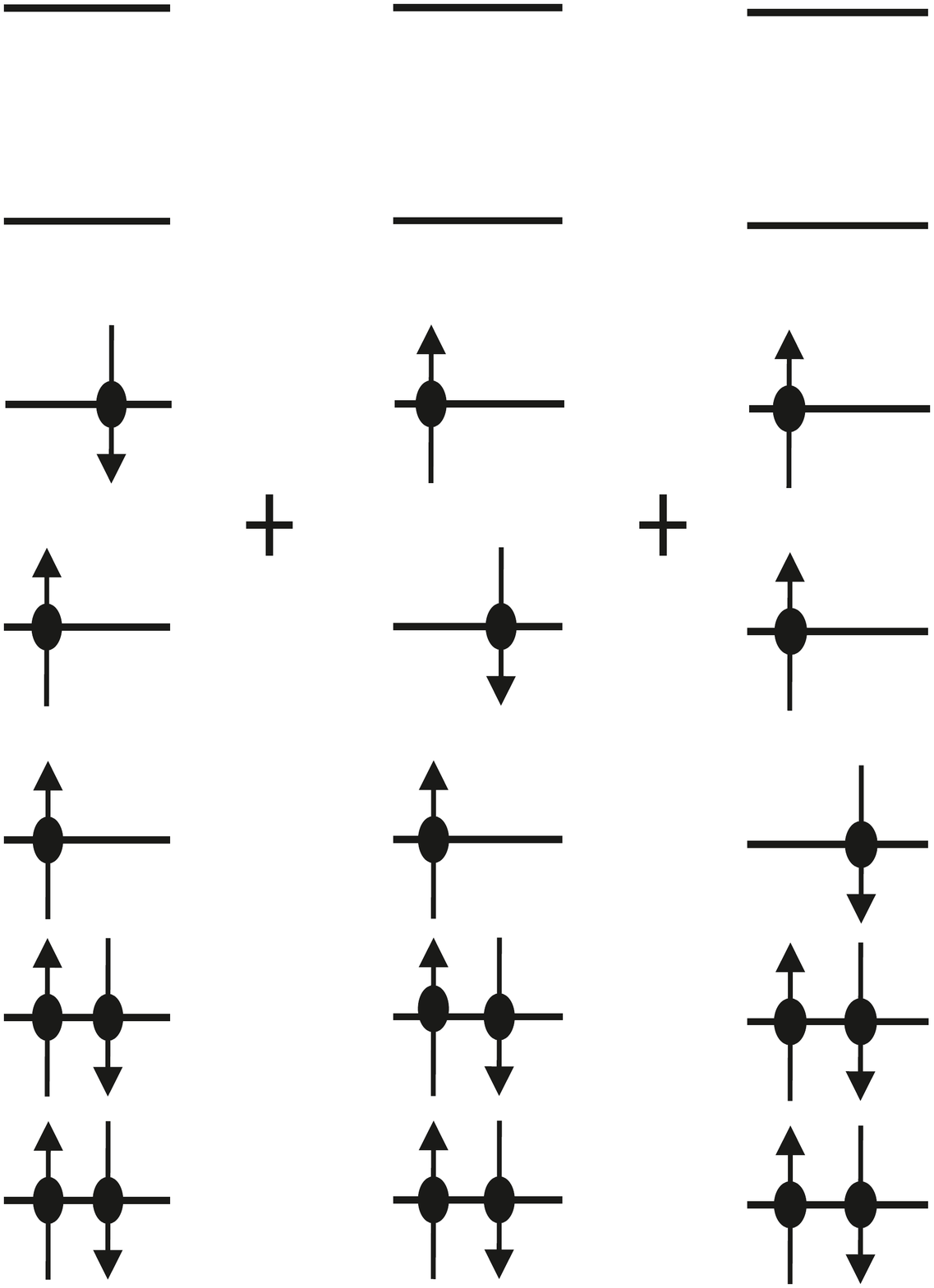} b)\includegraphics[width=0.2\textwidth]{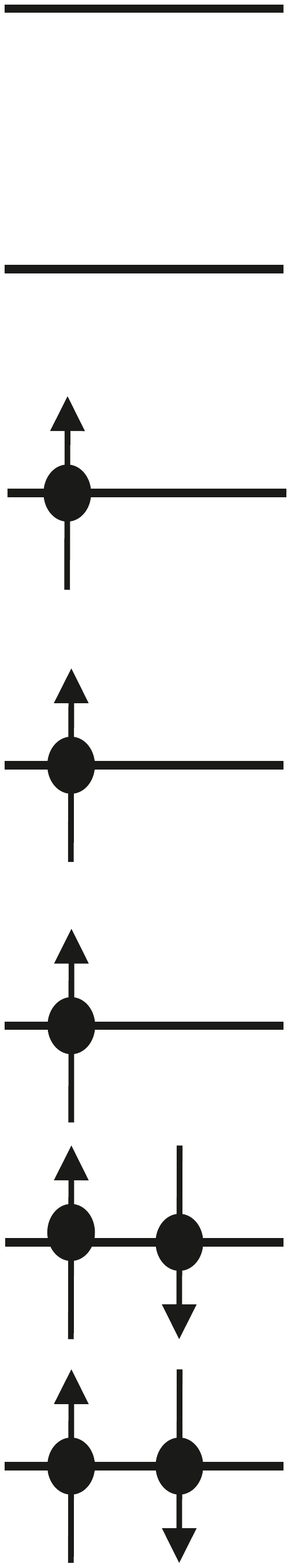}}
	\caption{Examples of low-energy eigenstates with the total spin $S=3/2$: a) $S_z=1/2$ and b)  $S_z=3/2$.}
	\label{Fig:Fig1}
\end{figure}

\subsubsection{Inelastic scattering cross section in the absence of magnetic field}

\begin{figure}[t]
	\centerline{\includegraphics[width=0.4\textwidth]{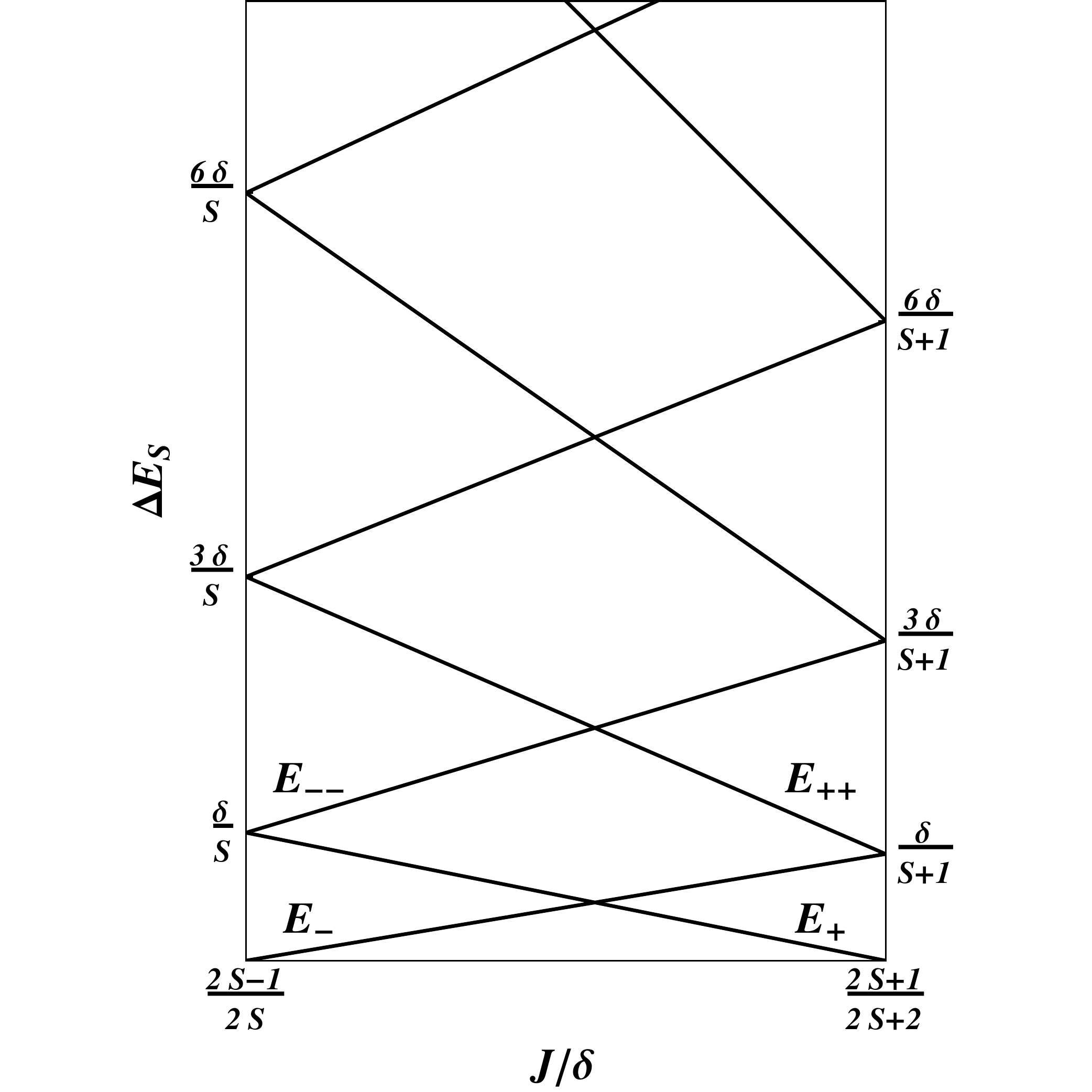}}
	\caption{Energies of the low-lying many-body eigenstates \eqref{eq:lles} as a function of $J/\delta$ for the case when the total spin in the ground state is equal to $S$. The ground state energy $E_S$ is set to zero.}
	\label{Fig:Fig2}
\end{figure}

Now let us consider the case of small electron energies $\varepsilon \ll \delta, J$. We assume that the quantum dot is in the regime of mesoscopic Stoner instability, $\delta,J \gg \delta-J$. Also we consider the case of low temperatures $T \lesssim \delta-J  \ll \delta, J$. 
For simplicity, we consider the case of equidistant single-particle spectrum. Afterwards we discuss the effect of fluctuations of single-particle levels. The minimal energy of the many-body state with the total spin $S$ is equal to
\begin{equation}
	E_S=(\delta-J) S^2- J S .
	\label{eq:lles}
\end{equation}
Here we omit the term proportional to the charging energy $E_c$ since we discuss the states with the same number of electrons. These many-body states consists of the three groups of levels: doubly occupied levels at the bottom, singly occupied levels in the middle, and empty levels at the top  (see Fig. \ref{Fig:Fig1}). Provided the exchange interaction is bounded to the following interval 
\begin{equation}
\frac{2S-1}{2S} < J/\delta < \frac{2S+1}{2S+2}, 
 \label{eq:int:S}
 \end{equation}
 the quantum dot has the total spin $S$ in the ground state. For $\delta-J\ll\delta, J$ its value is large, $S \approx \delta/[2(\delta-J)] \gg 1$. Interestingly, in this regime there are two low lying many-body excited states which corresponds to the states with the total spins $S+1$ and $S-1$. The gaps $E_\pm =E_{S\pm 1} - E_S$ between these excited states and the ground state is much smaller then the typical level spacing: $E_+ = (\delta-J)(2S+1)-J \leqslant \delta/S$ and $E_-=-(\delta-J)(2S-1)+J\leqslant \delta/(S+1)$. For the case of large total spin, $S\gg 1$, the gaps $E_+$ and $E_-$ are small in comparison with the mean single-particle level splitting, $E_\pm \ll \delta$.
The next many-body excited states with the total spins $S\pm 2$ have the gaps which lies in the following intervals,
$\delta/(S+1) \leqslant E_{++} \leqslant 3 \delta /S$ and $\delta/S \leqslant E_{--} \leqslant 3 \delta /(S+1)$ (see Fig. \ref{Fig:Fig2}). Assuming that temperature $T\lesssim \delta-J$ we neglect them. 

\begin{figure}[t]
	\centerline{\includegraphics[width=0.4\textwidth]{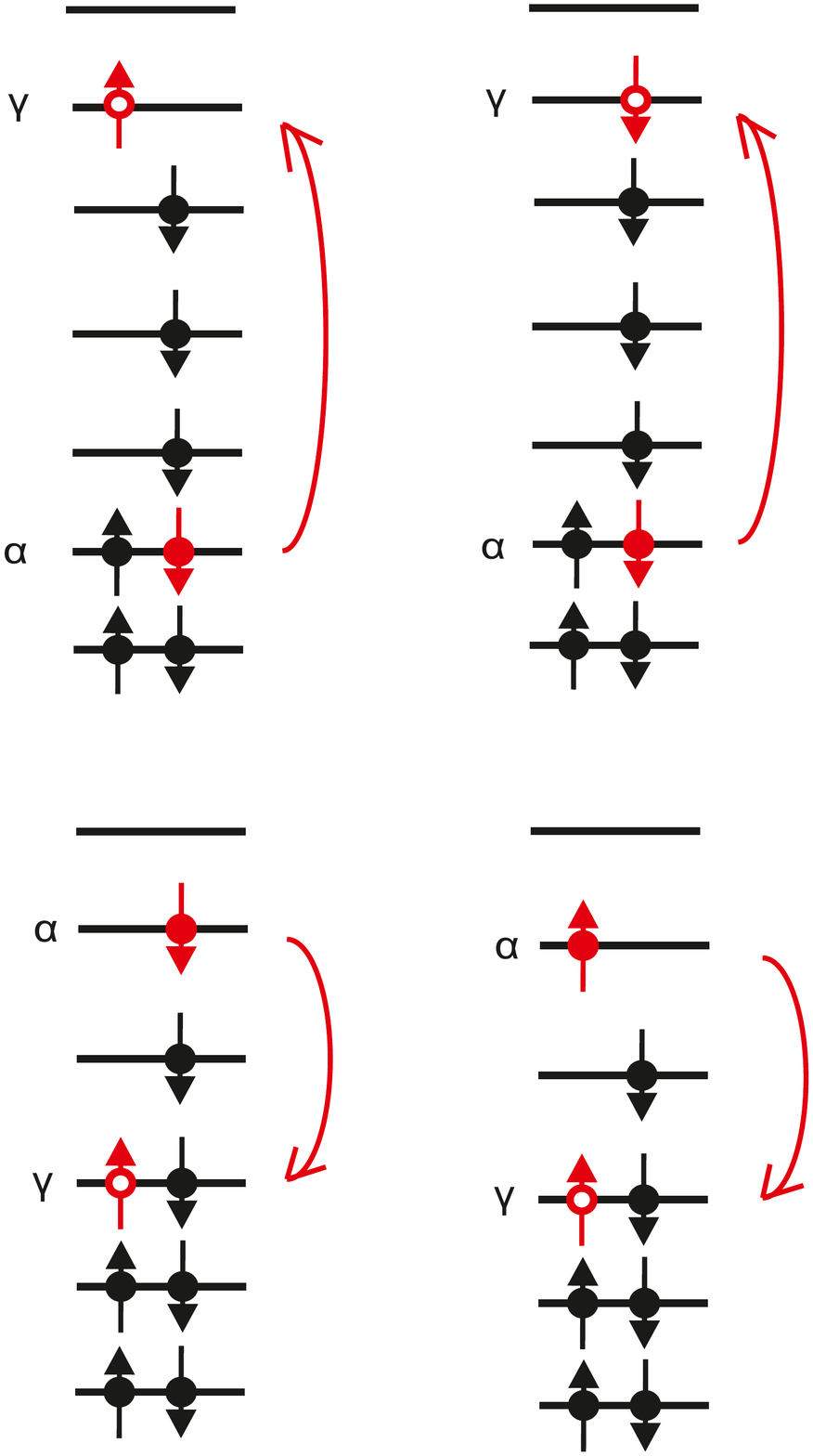}}
	\caption{(Color online) The sketch of inelastic transitions with (left column) and without (right column) spin-flip. The total spin increases (decreases) by one during the transition in the top (bottom) row (see text).}
	\label{Fig:Fig3}
\end{figure}

The operator $d^\dag_{\gamma \sigma^\prime} d_{\alpha\sigma}$ with $\alpha \ne \gamma$ has nonzero matrix elements between the many-body states with the same or shifted by one spin projection. Let us consider the ground state with the total spin $S$ and projection $M$. The state $d^\dag_{\gamma \sigma^\prime} d_{\alpha\sigma}|S,M\rangle$ will have the energy equal to $E_{S+1}$ if the level $\alpha$ is the highest doubly occupied one whereas the level $\gamma$ is the lowest empty one (see Fig. \ref{Fig:Fig3}).  The operator $d^\dag_{\alpha \bar\sigma} d_{\alpha\sigma}$ has nonzero matrix elements between the low lying many-body states with the same total spin. In this case the level $\alpha$ can be any among singly occupied levels those number is equal to $2S$. The corresponding matrix elements can be calculated in a standard way with the help of the Clebsch-Gordan coefficients (see e.g. Ref. [\onlinecite{LL3}]). The necessary matrix elements are summarized in Table \ref{Tab1}. Then for $T\lesssim \delta-J$ and $|\varepsilon|, \delta-J \ll \delta, J$ we find the following result for the inelastic scattering cross section:
\begin{align}
	{\mathcal{A}_{\rm inel}^\sigma(\varepsilon)} &= \frac{4\pi Q^2}{E_c^2} \Bigl \{ 
	\frac{(2S+1)(S+1)}{3}+\frac{1}{2}F(\varepsilon,E_-)\notag\\
	& + 
	\frac{2S+3}{2(2S+1)}F(\varepsilon,E_+)
	\Bigr \} ,
	\label{eq:final:inel-qd}
\end{align}
where we introduce the function 
\begin{equation}
F(\varepsilon,E) = \frac{2\cosh^2(\beta \varepsilon/2)}{\cosh(\beta \varepsilon) + \cosh (\beta E)} .
\label{eq:def:F}
\end{equation}

\begin{table*}[t]
\caption{Matrix elements between low-lying many-body states. The single particle states $\alpha$ and $\gamma$ are different, $\alpha\neq\gamma$ (see text and Fig. \ref{Fig:Fig3}).}
\begin{tabular}{c|c}
\hline
$\langle S+1,m+1 |d_{\gamma \uparrow}^\dagger d_{\alpha \downarrow}|S,m \rangle  =
\langle S+1,m+1 |(|S,m \rangle|1,1\rangle)=\frac{\sqrt{(S+m+2)(S+m+1)}}{\sqrt{(2S+1)(2S+2)
}}$
& $\sum\limits_{m=-S}^S \bigl |\langle S+1,m+1 |d_{\gamma \uparrow}^\dagger d_{\alpha \downarrow}|S,m \rangle \bigr |^2 = \frac{2S+3}{3}$ 
\\
$\langle S-1,m+1 |d_{\gamma \uparrow}^\dagger d_{\alpha \downarrow}|S,m \rangle =
\overline{\langle S,m |(|S-1,m+1 \rangle|1,-1\rangle)} = \frac{\sqrt{(S-m)(S-m-1)}}{\sqrt{(2S)(2S-1)}}$  
&
$\sum\limits_{m=-S}^S \bigl |\langle S-1,m+1 |d_{\gamma \uparrow}^\dagger d_{\alpha \downarrow}|S,m \rangle \bigr |^2 = \frac{2S+1}{3}$
\\
$\langle S+1,m |d_{\gamma \downarrow}^\dagger d_{\alpha \downarrow}|S,m \rangle =
\frac{1}{\sqrt{2}}\langle S+1,m |(|S,m \rangle|1,0\rangle)
=\frac{\sqrt{(S+m+1)(S-m+1)}}{\sqrt{(2S+1)(2S+2)}}$
&
$\sum\limits_{m=-S}^S \bigl |\langle S+1,m |d_{\gamma \downarrow}^\dagger d_{\alpha \downarrow}|S,m \rangle \bigr |^2 = \frac{2S+3}{6}
$
 \\
$\langle S-1,m |d_{\gamma \downarrow}^\dagger d_{\alpha \downarrow}|S,m \rangle =\frac{1}{\sqrt{2}}\overline{\langle S,m |(|S-1,m \rangle|1,0\rangle)} =\frac{\sqrt{(S+m)(S-m)}}{\sqrt{(2S)(2S-1)}}$ 
& 
$\sum\limits_{m=-S}^S \bigl |\langle S-1,m |d_{\gamma \downarrow}^\dagger d_{\alpha \downarrow}|S,m \rangle \bigr |^2 = \frac{2S+1}{6}$ \\
$\langle S,m+1 |d_{\alpha \uparrow}^\dagger d_{\alpha \downarrow}|S,m \rangle =\frac{1}{2S}\langle S,m+1 |S^+ |S,m \rangle =\frac{\sqrt{(S-m)(S-m+1)}}{2S}$ 
& $ \sum\limits_{m=-S}^S \bigl |\langle S,m+1 |  d_{\alpha \uparrow}^\dagger d_{\alpha \downarrow}|S,m \rangle \bigr |^2 = \frac{(S+1)(2S+1)}{6S}$ \\
\hline
\end{tabular}
\label{Tab1}
\end{table*}

The first contribution in Eq. \eqref{eq:final:inel-qd} represents the elastic spin-flip scattering, the next two correspond to the inelastic scattering with and without spin-flip. We note that the contribution in Eq. \eqref{eq:final:inel-qd}  due to the elastic spin-flip scattering, $(2S+1)(S+1)/3$, is larger than the result for the magnetic impurity, $2S(S+1)/3$. It occurs due to additional correlations between tunneling amplitudes in the case of orthogonal ensemble ($\bm{\beta}=1$). 

Now let us consider the case of higher temperatures, $\delta\gg T \gg \delta-J$. Then many low energy excited states with the total spin $S\pm k$ with $k\lesssim \sqrt{T/(\delta-J)}$ contribute to the inelastic cross-section. For $\delta-J\lesssim T$ the summation over discrete values of $k$ can be substituted by an integration. Using the following result, 
\begin{gather}
\frac{\int dS (2S+1) f(S) e^{-\beta E_S}}{\int dS (2S+1) e^{-\beta E_S}} 
=f\left ({S}_g\right )+\frac{T}{\delta} f^\prime\left ({S}_g\right ) \notag \\
+ \frac{T}{4(\delta-J)} f^{\prime\prime}\left ({S}_g\right ) ,
\end{gather}
where ${S}_g = {J}/[{2(\delta-J)}]$ and $f(S)$ is a quadratic polynomial of $S$, we obtain the inelastic scattering cross section for $\delta\gg T\gg\delta-J$ at energies $|\varepsilon| \lesssim \delta$ as follows
\begin{gather}
{\mathcal{A}_{\rm inel}^\sigma(\varepsilon)}  = \frac{4 \pi Q^2}{E_c^2}\Biggl [  \frac{\delta(3\delta-2J)}{6(\delta-J)^2} + \frac{T}{3(\delta-J)}\Biggr ] .
\label{eq:A:heis:ht}
\end{gather}
We emphasize that the inelastic cross section becomes larger than one can expect for the case of magnetic impurity with the spin of the order of $\delta/[2(\delta-J)]$.

Above we assumed that the single-particle level spacing in the quantum dot is equidistant. In general, this is not the case. Below following Ref. [\onlinecite{BGK2012}] we shall take into account fluctuation of the single-particle levels near the Stoner instability, $\delta-J\ll\delta$.  For a given realization of the single-particle  levels the energy $E_{+}$ acquires a random correction $\Delta E_{2S}$: $E_{+} \to E_{+}+\Delta E_{2S}$. This random energy correction is due to fluctuations of single-particle energy in a strip with $2S$ levels in average. It can be estimated as $\Delta E_{2S} = \delta \Delta n_{2S}$ where $\Delta n_{2S}$ stands for fluctuation of the number of levels in the energy strip with $2S$ levels in average. Near the Stoner instability we obtain from the condition $E_{+}+\Delta E_{2S}=0$ that the spin in the ground state is given as
\begin{equation}
S = \frac{\delta}{2(\delta-J)} \bigl [ 1 - \Delta n_{2S} \bigr ] .
\label{eq:S:fl:1}
\end{equation}
It is well-known from the random matrix theory [\onlinecite{Mehta}] that for $S \gg 1$ the fluctuations of  $\Delta n_{2S}$ are Gaussian and 
\begin{equation}
\overline{\Delta n_{2S}}=0, \qquad \overline{\bigl (\Delta n_{2S}\bigr )^2} = \frac{2}{\bm{\beta}\pi^2} \Bigl ( \ln 2S + {\rm const} \Bigr ) .
\label{eq:S:fl:2}
\end{equation}
Then with the help of Eqs. \eqref{eq:S:fl:1} and \eqref{eq:S:fl:2}, for a function $f(S)$ which is the quadratic polynomial as in Eq. \eqref{eq:final:inel-qd} we find
\begin{equation}
\overline{f(S)} = f(\overline{S}) + \frac{\overline{S}^2}{\bm{\beta}\pi^2} \ln (2\overline{S}) f^{\prime\prime}(\overline{S})  ,
\label{eq:S:fl:3}
\end{equation}
where $\overline{S} = \delta/[2(\delta-J)]$.
Using Eq. \eqref{eq:S:fl:3} and averaging the functions $F(\varepsilon,E_++\delta \Delta n_{2S})$ and $F(\varepsilon,E_--\delta \Delta n_{2S})$ in Eq. \eqref{eq:final:inel-qd} over $\Delta n_S$ with Gaussian distribution \eqref{eq:S:fl:2}, we obtain the averaged inelastic scattering cross section for temperatures $T\lesssim \delta-J$ and energies $\varepsilon\ll \delta$:
 \begin{align}
	\overline{\mathcal{A}_{\rm inel}^\sigma(\varepsilon)} & =  \frac{4\pi Q^2}{E_c^2} \Biggl \{ 
	\frac{2\delta^2}{3(\delta-J)^2} \Bigl [ 1+\frac{2}{\pi^2}\ln\frac{\delta}{\delta-J}\Bigr ] \notag \\
	 & +\mathcal{F}\left (\varepsilon,2(\delta-J),\frac{\delta}{\pi} \sqrt{2 \ln[\delta/(\delta-J)]}\right )
		\Biggr \} .
	\label{eq:final:inel-qd-n}
\end{align}
Here we neglect subleading terms in comparison with the largest one in the first line of Eq. \eqref{eq:final:inel-qd-n} which corresponds to elastic spin-flip contribution. The function $\mathcal{F}(x,y,z)$ is defined as follows
\begin{equation}
\mathcal{F}(x,y,z) = 1  + \frac{1}{2} \erf\left (\frac{x-y}{z}\right )  
- \frac{1}{2} \erf\left (\frac{x+y}{z}\right ) ,
\end{equation}
where $\erf(z) = (2/\sqrt{\pi})\int_0^z dt \exp(-t^2)$ denotes the error function. The result \eqref{eq:final:inel-qd-n} is valid under the following assumptions
\begin{equation}
(1-J/\delta)^2 \ll ({2}/{\pi^2}) \ln \bigl [\delta/(\delta-J)\bigr ] \ll 1 .
\label{eq:ineq:S}
\end{equation}
This restricts the value of the total spin in the ground state 
to the interval $2\lesssim \overline{S} \lesssim 70$. The right inequality in Eq. \eqref{eq:ineq:S} guaranties that fluctuations of $S$ are small and Gaussian. For $\overline{S} \gg (1/2) \exp(\pi^2/2)$ fluctuations of the total spin becomes non-gaussian (see Refs. [\onlinecite{Sharafutdinov2014a,*Sharafutdinov2014b}]). The left inequality in Eq. \eqref{eq:ineq:S} guaranties that the effective temperature $T_{\rm eff} \sim (\delta/\pi) \sqrt{2 \ln [\delta/(\delta-J)]}$ induced by fluctuations and smearing the steps at $\varepsilon = \pm 2(\delta-J)$ is larger than the temperature, $T_{\rm eff} \gg \delta-J \gtrsim T$. We note that the effective temperature is low in comparison with the mean level spacing, $T_{\rm eff} \ll \delta$. All in all, fluctuations of the single-particle levels enhances the elastic spin-flip contribution (similarly to enhancement of spin susceptibility [\onlinecite{KAA,BGK2012,Sharafutdinov2014a,*Sharafutdinov2014b}]) and smear the steps in the inelastic spin-flip and non-spin-flip contributions.

\subsubsection{Inelastic scattering cross section in the presence of magnetic field}

Now we consider the behavior of inelastic cross section in the presence of magnetic field $B$. We assume that in addition to the Zeeman splitting this magnetic field produces the orbital effect
and breaks the time reversal symmetry. Then the parameter $\bm{\beta}$ becomes equal to $2$, $\bm{\beta}=2$. We consider the case of Zeeman splitting which is strong in comparison with temperature but small with respect to $\delta$, $\delta \gg b= \mu_B g_L B\gg T$. Then the degeneracy of the low lying states with the total spin $S$ is removed. The lowest energy state with the total spin $S$ corresponds to the maximal total spin projection along the magnetic field $S_z=S$ (we assume $B>0$). 
The energies of these states become
\begin{equation}
E_S(B)=(\delta-J) S^2-J S-b S .
\end{equation}
Hence, in the presence of Zeeman splitting the total spin in the ground state
is equal to $S\approx{(\delta+b)}/{[2(\delta-J)]}$ for $\delta-J\ll\delta$. 

The absence of degeneracy with respect to the total spin projection makes the elastic spin-flip contribution to the inelastic cross section to be exponentially small in parameter $\beta b \gg 1$. The same holds for the contribution due to inelastic scattering without spin-flip. Thus the main contribution to the inelastic scattering cross section comes from inelastic spin-flip scattering:
\begin{equation}
 {\mathcal{A}_{\rm inel}^\sigma(\varepsilon)}  = \frac{4 \pi Q^2}{E_c^2} \sum_{\alpha\neq\gamma} {\sum_{f\neq  i}}^\prime  p_i \frac{[1+e^{-\beta\varepsilon}]\bigl | \langle i | d^\dag_{\gamma,-\sigma} d_{\alpha\sigma}  | f\rangle  \bigr |^2}{1+e^{-\beta(\varepsilon+E_i-E_f)}} .
 \label{eq:A:21}
\end{equation}
Here the prime sign indicates that the summation is over the low-energy many-body states  $i$ and $f$ which characterized by the total spin $S$ and the maximal total spin projection along the magnetic field, $S_z=S$ (see Fig. \ref{Fig:Fig1}b). The gaps between the ground state with the total spin $S$ and the lowest many-body excited states, $E_\pm(B) = E_\pm \mp b$, can be bounded from above:
$\max E_+(B) \leqslant (\delta+b)/S$  and $\max E_-(B) \leqslant (\delta+b)/(S+1)$. We note that for $S\gg 1$ the energy scale $(\delta+b)/S \approx 2(\delta-J) \ll \delta$. To calculate the matrix elements in \eqref{eq:A:21} one needs to take into account that the single-particle level $\alpha$
should be the highest doubly occupied level and $\gamma$ should be the lowest unoccupied level or 
vice versa (see transition in the left lower corner of Fig. \ref{Fig:Fig3}). Using the results for the matrix elements from the table \ref{Tab1} we find the following result for the inelastic cross section at $|\varepsilon| \ll \delta$ and $T \lesssim (\delta-J) \ll b$:
\begin{equation}
 {\mathcal{A}_{\rm inel}^\sigma(\varepsilon)} =\frac{4\pi Q^2}{E_c^2} \Bigl [ 1-n_F\bigl (\varepsilon-E_\sigma(B)\bigr )+n_F\bigl (\varepsilon+E_{\bar{\sigma}}(B)\bigr ) \Bigr ] .
\label{eq:r:t6}
\end{equation}
As one can check the result for ${\mathcal{A}_{\rm inel}^\sigma(\varepsilon)}$ at $B<0$ can be obtained from the result for ${\mathcal{A}_{\rm inel}^{\bar{\sigma}}(\varepsilon)}$ for $B>0$. 

It is instructive to compare the results for ${\mathcal{A}_{\rm inel}^\sigma(\varepsilon)}$
with and without magnetic field. At first, the magnetic field suppresses the elastic spin-flip contribution. Secondly, instead of four steps of height $1/2$ (in case of large spin $S\gg1$) at energies $E_\pm$ and $-E_\pm$ in the absence of magnetic field (see Eq. \eqref{eq:r:t6}), in the presence of the Zeeman splitting only two steps at $E_\sigma(B)$ and $-E_{\bar\sigma}(B)$ of height $1$ survive. We stress that, contrary to the case of magnetic impurity  the inelastic scattering cross section off the quantum dot  in the presence of the Zeeman splitting at energies $|\varepsilon| \gg E_\sigma(B)$ are not exponentially small in $\beta b \gg 1$. For energies $|\varepsilon|\ll E_\sigma(B)$ the inelastic scattering cross section is zero at $T=0$.

In the presence of fluctuations of the single-particle levels the energies $E_\sigma(B)$ become random,  
$E_\sigma(B) \to E_\sigma(B)  +\sigma \Delta E_{2S}$. As a consequence, the spin in the ground state becomes fluctuating:
\begin{equation}
S = \frac{1}{2(\delta-J)} \Bigl [ \delta+b - \delta \Delta n_{2S} \Bigr ] .
\end{equation}
Averaging the Fermi functions in Eq. \eqref{eq:r:t6} over $\Delta n_{2S}$ over the Gaussian distribution \eqref{eq:S:fl:2} with $\bm{\beta}=2$, we find the inelastic scattering cross section at $|\varepsilon| \ll \delta$ and $T \lesssim (\delta-J) \ll b\ll \delta$ as
  \begin{align}
 \overline{\mathcal{A}_{\rm inel}^\sigma(\varepsilon)} =\frac{4\pi Q^2}{E_c^2} \Biggl [ 1 & +
 \frac{1}{2} \erf \left ( \frac{\pi [\varepsilon-2(\delta-J)]}{\delta\sqrt{\ln[b/(\delta-J)]}}\right )
\notag \\
 & - \frac{1}{2} \erf \left ( \frac{\pi [\varepsilon+2(\delta-J)]}{\delta\sqrt{\ln[b/(\delta-J)]}}\right )
  \Biggr ] .
\label{eq:r:t6-n}
\end{align}
This result is valid provided the following inequality holds:
\begin{equation}
(\delta-J)^2 \ll \frac{1}{\pi^2} \ln \frac{b}{\delta-J} \ll 1 .
\label{eq:ineq:22}
\end{equation} 
The left inequality in Eq. \eqref{eq:ineq:22} implies that the effective temperature $T_{\rm eff} = (\delta/\pi)\sqrt{\ln[b/(\delta-J)]}$ smearing the steps in $\mathcal{A}_{\rm inel}^\sigma(\varepsilon)$  at $E_\sigma(B)$ and $-E_{\bar\sigma}(B)$ is not very low, $T_{\rm eff} \gg \delta-J \gtrsim T$. The right inequality in Eq. \eqref{eq:ineq:22} guaranties that fluctuations of the total spin remains Gaussian.

\subsection{Inelastic scattering cross section in the presence of strong spin-orbit coupling}
\label{Sec:MLQD:Ising}

In the previous section we demonstrate that the Zeeman splitting suppresses the elastic spin-flip scattering due to lifting the $2S+1$ degeneracy of the ground state. In this section we discuss another  
mechanism of suppression of the elastic spin-flip scattering on the quantum dot. We consider a quantum dot fabricated in 2D electron gas with strong spin-orbit coupling. Such quantum dot can be described by the universal Hamiltonian \eqref{eq:H:QD} in which the Heisenberg exchange is substituted by the Ising exchange: $J \bm{S}^2 \to J S_z^2$ [\onlinecite{Falko},\onlinecite{AlhassidRupp}]. In this case the statistics of single particle levels is described by the unitary symmetry ensemble (class A) with $\bm{\beta}=2$.

The low energy many-body states correspond to the total spin $S$ and the maximal or minimal spin projection $S_z=\pm S$. The energies of these states are equal to
\begin{equation}
E_S = (\delta-J) S^2  .
\end{equation}  
Therefore the total spin in the ground state is equal to 0 ($1/2$) in case of even (odd) number of electrons. 

For the even number of electrons, since $S=0$, the elastic spin-flip scattering vanishes. For $T\lesssim \delta-J$ and $|\varepsilon|\lesssim \delta$, the only contribution to ${\mathcal{A}_{\rm inel}^\sigma(\varepsilon)}$ remains due to the inelastic spin-flip:
\begin{equation}
{\mathcal{A}_{\rm inel}^{\sigma,e}(\varepsilon)}  = \frac{4 \pi Q^2}{E_c^2}
\Bigl [ 1-n_F\bigl (\varepsilon-\Delta_e\bigr )+n_F\bigl (\varepsilon+\Delta_e\bigr ) \Bigr ] .
\label{eq:d1:even}
\end{equation}
Here $\Delta_e = \delta-J$ stands for the gap between the ground state with $S=S_z=0$ and the states with $S=1$ and $S_z=\pm 1$.

In the case of the odd number of electrons, since $S=1/2$, the ground state is doubly degenerate. Then the elastic spin-flip scattering is the same as for the magnetic impurity with spin $1/2$. In addition, inelastic spin-flip contributes to the inelastic cross section. Then at $T\lesssim \delta-J$ and $|\varepsilon|\lesssim \delta$ we find
\begin{equation}
{\mathcal{A}_{\rm inel}^{\sigma,o}(\varepsilon)}  = \frac{2 \pi Q^2}{E_c^2}
\Bigl [1 + 1-n_F\bigl (\varepsilon-\Delta_o\bigr )+n_F\bigl (\varepsilon+\Delta_o\bigr ) \Bigr ] .
\label{eq:d1:odd}
\end{equation}
Here $\Delta_o = 2(\delta-J)$ denotes the gap between the ground state with $S=1/2$ and the states with $S=1$ and $S_z=\pm 1$.

We note that the inelastic scattering rate at energies $|\varepsilon| \gg \delta-J$ is independent of electron parity in the quantum dot, 
\begin{equation}
{\mathcal{A}_{\rm inel}^{\sigma,e}(\varepsilon)}={\mathcal{A}_{\rm inel}^{\sigma,o}(\varepsilon)} = \frac{4 \pi Q^2}{E_c^2}
 .
 \label{A:eq:L}
\end{equation}

In the case of temperatures $\delta \gg T \gg \delta-J$, the low-lying many-body states with the total spin $S\lesssim \sqrt{T/(\delta-J)}$ contribute to the inelastic cross section for $|\varepsilon| \lesssim \delta$. Similar to low temperatures, the dominant contribution is due to inelastic spin-flip. Then we obtain
\begin{align}
{\mathcal{A}_{\rm inel}^{\sigma}(\varepsilon)} & = \frac{8 \pi Q^2}{E_c^2} \left [\sum_{S_z=-\infty}^\infty 
e^{\beta(J-\delta)S_z^2} \right ]^{-1} \sum_{S_z=-\infty}^\infty 
e^{\beta(J-\delta)S_z^2}\notag \\
& \times  F\bigl (\varepsilon,(\delta-J)(2S_z+1)\bigr )  = \frac{8 \pi Q^2}{E_c^2} .
\end{align}
We note that the the inelastic cross section for $|\varepsilon|\sim \delta$ at $\delta \gg T \gg \delta-J$ is twice larger than at $T\ll \delta-J$. This difference stems from the following. At high temperatures, $\delta \gg T \gg \delta-J$, the following four combinations of initial and final states  contribute to the inelastic spin-flip cross section (see Eq. \eqref{eq:r:t4a}): (i) $| i \rangle = |S, S\rangle$ and $| f\rangle = |S-1, S -1\rangle$; (ii) $| i\rangle = |S, -S\rangle$ and $| f\rangle = |S+1,-S -1\rangle$; (iii) $| i\rangle = |S+1, S+1\rangle$ and $| f\rangle = |S, S\rangle$; (iv) $| i\rangle = |S-1, -S+1\rangle$ and $| f\rangle = |S, -S\rangle$. In the case of low temperatures, $T\ll \delta-J$, when the state with the lowest spin ($S=0$ or $S=1/2$) contribute only, the transitions (i) and (iv) are not possible. 

In case of the even number of electrons the gap $\Delta_e$ is determined by the difference in the level spacing between the lowest singly occupied and the highest doubly occupied levels and the exchange energy. As it is well-known, the level spacing strongly fluctuates and its distribution can be well approximated by the Wigner Surmise (see Ref. [\onlinecite{Mehta}]). The typical scale of this distribution is given by the mean level spacing. Qualitatively, averaging  of Eq. \eqref{eq:d1:even} over distribution of $\Delta_e$ results in the same form of the dependence on energy but with effective temperature proportional to the mean level spacing.  Similar results one obtains after averaging of Eq. \eqref{eq:d1:odd}. The inelastic cross section at temperatures $\delta \gg T \gg \delta-J$ is robust with respect to fluctuations since it is independent of particular properties of the single-particle spectrum.

\section{Discussions and conclusions}

Our results for the inelastic scattering cross section of an electron off the quantum dot at low temperatures allow us to estimate corresponding contribution to the dephasing rate. Assuming a finite concentration $n_s$ of quantum dots we introduce the inelastic scattering rate for an electron as follows 
\begin{equation}
	 {\tau^{-1}_{\rm inel}(\varepsilon,T)} = \frac{n_s}{\nu} \sum_{\sigma=\pm} {\mathcal{A}_{\rm inel}^\sigma(\varepsilon)} .
\end{equation}
We remind that $\nu$ denotes the average density of states per spin for electrons in the electron liquid surrounding quantum dots. We note that in Refs. [\onlinecite{Micklitz2006}, \onlinecite{Kettemann2007}] the inelastic rate at finite temperature has been related directly to the difference between the imaginary part of the T-matrix and the diagonal element of its square. Although this is correct for the case of zero temperature
at finite temperature it is not the case in general (see discussion after Eq. \eqref{eq:Nch}).  In our case, by definition, the quantity ${\mathcal{A}_{\rm inel}^\sigma(\varepsilon)}$ includes the inelastic processes only.

Using the fact that the quantity $1/{\tau_{\rm inel}(\varepsilon,T)}$ represents the self-energy for the electron pair propagator (cooperon) one can estimate the dephasing time $\tau_\phi(T)$ entering the expression for the weak-localization correction to the conductivity [\onlinecite{Micklitz2006}]. The concrete expression depends on dimensionality. Having in mind experiments of Refs. [\onlinecite{Kuntsevich1,*Kuntsevich2}] we restrict our discussion to two dimensions, $d=2$. In this case, one can obtain [\onlinecite{Micklitz2006}]:
\begin{equation}
\tau^{-1}_\phi(T) = \exp \left [ \int d \varepsilon \, n_F^\prime(\varepsilon) \ln \tau_{\rm inel}(\varepsilon,T)\right ] .
\label{eq:tauphi}
\end{equation}
We mention that the above estimate for $\tau_\phi(T)$ is based on independent treatment of the inelastic scattering off quantum impurities and elastic disorder scattering. As discussed in Ref. [\onlinecite{Micklitz2006}], such simplified approach is valid for $d=2$ under the following assumptions: (i) the system is very good metal: the conductance $g \gg (\nu J_s)^{-3}$; (ii) the density of quantum impurities is not large, $n_s \ll \nu T_K$. For our problem the characteristic exchange interaction between electrons and quantum impurities (quantum dots) can be estimated as $\nu J_s = Q/E_c$ (see Sec. \ref{Sec:Form:1LQD}). The Kondo temperature $T_K$ is given by the standard expression, $T_K \sim E_c \exp(-1/\nu J_s)$.  

We start the discussion of $\tau_\phi(T)$ from the case of isotropic exchange interaction on the quantum dot. Near the Stoner instability, $\delta-J\ll \delta$, at temperatures $T\ll \delta-J$, the inelastic scattering rate ${\mathcal{A}_{\rm inel}^\sigma(\varepsilon)}$ is given by Eq. \eqref{eq:final:inel-qd}. Performing expansion in small energy-dependent terms we find from Eq. \eqref{eq:tauphi}
\begin{align}
\tau^{-1}_\phi(T) & = \frac{8 \pi n_s Q^2}{\nu E_c^2} \Biggl [ \frac{(S+1)(2S+1)}{3} + \beta E_- e^{-\beta E_-}
\notag \\
& +\frac{2S+3}{2S+1} \beta E_+ e^{-\beta E_+}
\Biggr ]\label{eq:tauphi1}
\end{align}
There is a weak temperature dependence of the dephasing rate due to possibility of inelastic scattering which involves transitions to the lowest many-body levels of the quantum dot. Also we emphasize that 
the elastic spin-flip contribution to the dephasing rate is different from a standard one for a magnetic impurity which is proportional to $S(S+1)/3$. We repeat that it occurs due to additional correlations between tunneling amplitudes for transitions to different levels of the quantum dot. Using Eq.  \eqref{eq:A:heis:ht} we obtain at higher temperatures $\delta-J \ll T \ll \delta$:
\begin{equation}
\tau^{-1}_\phi(T) = \frac{8 \pi n_s Q^2}{\nu E_c^2} \Biggl [  \frac{\delta(3\delta-2J)}{6(\delta-J)^2} + \frac{T}{3(\delta-J)}\Biggr ]  .
\label{eq:tauphi2}
\end{equation}
We mention that in fact both estimates \eqref{eq:tauphi1} and \eqref{eq:tauphi2} hold for dimension $d=3$ as well. In the presence of Zeeman splitting, the elastic spin-flip is suppressed. Then using Eq. \eqref{eq:r:t6}, we find the following estimate for the dephasing rate at low temperatures $T\ll \delta-J$ and moderate magnetic fields, $\delta \gg b \gg \delta -J$:
\begin{equation}
\tau^{-1}_{\phi}(T) = \frac{4 \pi e^2 n_s Q^2}{\nu E_c^2} \left [e^{-\beta E_+(B)}+ e^{- \beta E_-(B)} \right ].
\label{eq:tauphi3}
\end{equation}
We note that the dephasing rate in this case is exponentially small in temperature, $\tau^{-1}_{\phi}(T) \sim \exp(-2 \beta (\delta-J))$.

In the case of Ising exchange interaction on the quantum dot $\tau_\phi(T)$ depends on the parity of the number of electrons  at low temperatures $T\ll \delta-J$. Using Eqs. \eqref{eq:d1:even} and \eqref{eq:d1:odd}, we obtain
\begin{equation}
\tau^{-1}_{\phi,e}(T) = \frac{8 \pi n_s Q^2}{\nu E_c^2} e^{-\beta \Delta_e}
\end{equation}
for the even number of electrons, and
\begin{equation}
\tau^{-1}_{\phi,o}(T) = \frac{4 \pi n_s Q^2}{\nu E_c^2} \left [1 + \pi e^{-\beta \Delta_o} \right ]
\end{equation}
for the odd number of electrons. At higher temperatures, $\delta \gg T \gg \delta-J$, the dephasing time becomes insensitive to the parity of the number of electrons:
\begin{equation}
\tau^{-1}_{\phi}(T) = \frac{16 \pi n_s Q^2}{\nu E_c^2}  .
\end{equation}
Thus, the dephasing time for the temperature range, $\delta \gg T \gg \delta-J$, due to scattering off the quantum dot with Ising exchange is similar to the magnetic impurity.

We note that our approach completely ignores the effect of electron reservoir on the quantum dot. First of all, the coupling to the reservoir results in the broadening ($\Gamma$) of the single-particle levels which is of the order of $g_T\delta$. It can be neglected provided temperatures are not too low, $T\gg g_T \delta$. 
Secondly, due to coupling to the reservoir, the probabilities $p_i$ of many-body states of the quantum dot can become nonequilibrium, i.e. very different from the Gibbs form. However, in the case of slow escape rate (which is of the order of $\Gamma \sim g_T \delta$) in comparison with the intrinsic inelastic rate $1/\tau_{ee}$ due to electron-electron interaction inside the quantum dot, this nonequilibrium effect can be neglected. For the quantum dot of size larger than the mean free path $l$, the intrinsic inelastic rate can be estimated as [\onlinecite{Aronov},\onlinecite{Blanter}]: $1/\tau_{ee} \sim T^2 /(g_{\rm Th}^2 \delta)$. The condition $\Gamma\ll 1/\tau_{ee}$ results in the following restriction on temperatures at which our assumption of the equilibrium for the quantum dot holds:
\begin{equation}
T\gg \delta (g_T g_{\rm Th}^2)^{1/2} .
\end{equation}
Since in this work we study temperatures below $\delta-J$, the following condition for the tunneling conductances (or for proximity to the Stoner instability) emerges:
\begin{equation}
g_T \ll \frac{1}{g_{\rm Th}^2} \left (\frac{\delta-J}{\delta}\right )^{2} .
\end{equation}

Also our approach neglects the effect of the reservoir on the dynamics of the total spin in the quantum dot. In particular, we neglect the renormalization of the value of the total spin due to coupling to the reservoir. Using adiabatic approximation for the large total spin of the quantum dot near the Stoner instability [\onlinecite{Saha2012}], one can demonstrate that the spin moves diffusively on the Bloch sphere with the diffusive constant proportional to the tunneling coupling $Q$ [\onlinecite{Shnirman2015},\onlinecite{Shnirman2016}]. 

Recent experiments [\onlinecite{Kuntsevich1,*Kuntsevich2}] give an evidence which may be interpreted as formation of local spin droplets in 2D disordered electron liquid at low temperatures. As known [\onlinecite{Finkelstein1990}], at low temperatures 2D disordered electron liquid tends to the Stoner instability such that the renormalized Fermi-liquid interaction constant in the triplet channel tends to $-1$, $F_0^\sigma \approx -1$. The creation of spin droplets with a finite spin $S_g = 1/[2(1+F_0^\sigma)] \gg 1$ near the Stoner instability in disordered electron liquid due to fluctuations in the triplet (spin) channel has been predicted in Ref. [\onlinecite{Narozhny2000}]. The Pauli spin susceptibility $\chi\sim \nu/(1+F_0^\sigma)$ dominates at high temperatures. Due to presence of spin droplets one expects that the spin susceptibility is dominated by the Curie-like temperature dependence, $\chi \sim n_s S_g^2/T$ at low temperatures $T\ll T_* = n_s^{fl} S_g /\nu$. Here $n_s^{fl}$ denotes the density of spin droplets. We note that in the experiments [\onlinecite{Kuntsevich1,*Kuntsevich2}] the spin susceptibility behaves as $\chi\sim T^{-2}$ suggesting strong temperature dependence of the droplet density $n_s^{fl}$. 
 The electron scattering off such spin droplets results in the following contribution to the dephasing rate: $n_s^{fl} S_g^2/(\nu g)$ where $g$ is the conductance of 2D disordered electron liquid [\onlinecite{Narozhny2000}]. Comparing this contribution with the standard dephasing rate due to electron-electron interaction in the triplet channel, $T/[g(1+F_0^\sigma)]$ [\onlinecite{AAbook}], one finds that the dephasing rate should saturate below the same crossover temperature $T_*$. Thus, in the presence of such spin droplets the Curie-like temperature dependence of the spin susceptibility should be accompanied by the temperature independent dephasing time. In contrast, in the experiments [\onlinecite{Kuntsevich1,*Kuntsevich2}] the strong temperature dependence of the spin susceptibility has been observed together with linear in temperature dephasing rate. 

Let us now assume that there are some electron puddles in 2D electron liquid. Then in such puddle some number of electrons can be localized. We model such a droplet of size $L_d \gg l \gg \lambda_F$ by a quantum dot with the Heisenberg exchange interaction $J=-F_0^\sigma \delta$. Here $\lambda_F$  denotes the Fermi wavelength. 
Then at temperatures  $T\ll T_\star = n_s S_g /\nu$, where, we remind, $S_g \approx J/[2(\delta-J)] = 1/[2(1+F_0^\sigma) \gg 1$ is the total spin of the droplet, one expects that the spin susceptibility is dominated by the Curie-like temperature dependence, $\chi \sim n_s S_g^2/T$. Using Eq. \eqref{eq:tauphi2} as an estimate for the contribution to the dephasing rate due to scattering off spin droplets we find that it dominates over the linear in $T$ contribution at temperatures $T\ll T_s \sim \eta^2 T_\star$ where the parameter $\eta = (Q/E_c) \sqrt{g} \sim g_{\rm ch} l/(r_s L_d \sqrt{g})$. Provided the interaction parameter $r_s \sim \lambda_F/a_B \sim 1$, where $a_B$ stands for the Bohr radius, and $g\gtrsim 1$, we find $\eta \ll 1$. Here we use the tunneling conductance per channel is small, $g_{\rm ch} \ll 1$. Thus, in the temperature range $T_s \ll T \ll T_\star$ we expect the Curie-like temperature dependence of the spin susceptibility but the conventional, linear in $T$, dephasing rate. The contribution to the dephasing rate due to electron scattering off electron puddles with a finite spin will dominate the contribution from scattering off spin droplets emerging due to fluctuations in the triplet channel [\onlinecite{Narozhny2000}] if the following condition holds $n_s^{fl} \ll \eta^2 n_s$ or, equivalently, $T_*\ll T_s$. Since the density of the spin droplets $n_s$ cannot be larger then $1/L_d^2$, the saturation temperature is below the mean level spacing on the quantum dot, $T_s\ll \delta$ provided $\eta\ll 1/\sqrt{S_g}$. To validate our proposition of two characteristic temperatures $T_s$ and $T_\star$ more detailed data on the structure and properties of spin droplets are needed. 
 
To summarize we studied the electron scattering off a quantum dot with large charging and 
exchange energies. We consider the scattering due to tunneling between electron liquid and the quantum dot. Under the following assumptions: (i) the quantum dot is in regime of a strong Coulomb blockade with integer number of electrons (Coulomb valley); (ii) the quantum dot is near the Stoner instability; we compute that inelastic cross section in the forth order in the tunneling amplitudes. We have analyzed in detail the behavior of the inelastic cross section at low temperatures and energies, $T, |\varepsilon| \ll \delta$, for three cases: the quantum dot with Heisenberg exchange without and with Zeeman splitting, and the quantum dot with Ising exchange. Using our results for the inelastic cross section we estimate the corresponding contributions to the electron dephasing rate. We use our results to estimate the temperature below which the dephasing time due to scattering off spin droplets in 2D disordered electron liquid should saturate. In agreement with the experiments we found that it is well below the temperature scale below which the spin susceptibility is expected to demonstrate Curie-like behavior.

\begin{acknowledgments}
We acknowledge useful discussions with Y. Gefen and M. Goldstein. We are grateful to A.Yu. Kuntsevich and V.M. Pudalov for numerous discussions of their experimental data. The research was partially funded by RFBR under the grant No. 15-52-06005. 
\end{acknowledgments}

\appendix

\section{Evaluation of the Green's function for electrons on the quantum dot to the second order in tunneling}
\label{App:A}

In this appendix we present some details of derivation of the results \eqref{eq:4}. First of all, it is convenient to rewrite the definition of the Green's function in the imaginary time $\tau>0$ (see Eq. \eqref{eq:3}) in the interaction representation as
\begin{align}
	\mathcal{G}_{\alpha\sigma;\beta\sigma^\prime}(\tau) = & - \frac{1}{\mathcal{Z}} \Tr \Bigl [ e^{-\tau H_0} U(\tau) d^\dag_{\beta\sigma^\prime} e^{-(\beta-\tau) H_{0}} \notag \\
&	\times U(\beta-\tau) d_{\alpha\sigma} \Bigr ]  ,
\label{eq:app1}
\end{align}
where $H_0 = H_{QD}+H_{R}$ and
\begin{equation}
U(\tau)  = \mathcal{T}_\tau \exp \left (- \int_0^\tau d\tau^\prime e^{\tau^\prime H_0} H_T e^{-\tau^\prime H_0} \right ) .
\end{equation}
Next we expand $U(\tau)$ to the second order in the tunneling Hamiltonian,
\begin{align}
	U(\tau)  \simeq & 1-   \int_0^\tau d\tau^\prime e^{\tau^\prime H_0} H_T e^{-\tau^\prime H_0}	+\int_0^\tau d\tau^\prime e^{\tau^\prime H_0} H_T  \notag \\
& \times e^{-\tau^\prime H_0}\int_0^{\tau^\prime} d \tau^{\prime\prime} e^{\tau^{\prime\prime} H_0} H_T e^{-\tau^{\prime\prime} H_0} ,
\end{align}
substitute this result into the expression \eqref{eq:app1} for Green's function, and take the trace over the reservoir degrees of freedom with the help of identity 
\begin{gather}
  \frac{\Tr \Bigr [ \mathcal{T}_\tau {a}_{k\sigma}(\tau) {a}^\dagger_{k^\prime\sigma^\prime}(0) e^{-\beta H_{R}} \Bigl ]}{\Tr e^{-\beta H_{R}}} = \delta_{\bm{k},\bm{k^\prime}}\delta_{\sigma,\sigma^\prime} e^{-\tau \varepsilon_{k\sigma}}\notag \\
\times  \Bigl[ \theta(\tau) - n_F(\varepsilon_{k\sigma}) \Bigr] ,
\end{gather}
where $\theta(\tau)$ stands for the Heaviside step function. Then we find,
\begin{widetext}
\begin{align}
	\mathcal{G}_{\alpha\sigma;\beta\sigma}(\tau)  = & \frac{1}{Z} \sum_{\gamma,\eta;\sigma^\prime}\int \limits_{-\infty}^\infty dE \, Q_{\gamma\eta}^{\sigma^\prime}(E) \notag \\
	\times &  \Biggl \{ \int\limits_0^{\beta-\tau} d\tau_1 \int\limits_0^\tau d\tau_2 \, \Bigr [ n_F(E) e^{(\tau+\tau_2-\tau_1)E} \Tr \Bigl ( 
	 d^\dag_{\gamma\sigma^\prime}e^{-\tau_1 H_{QD}} d^\dag_{\beta\sigma} e^{-(\beta-\tau-\tau_2)H_{QD}} d_{\eta\sigma^\prime} e^{-\tau_2 H_{QD}} d_{\alpha\sigma} e^{-(\tau-\tau_1) H_{QD}} \Bigr ) 
	 \notag \\
	 & + [1-n_F(E)] e^{-(\tau+\tau_2-\tau_1)E} \Tr \Bigl ( 
	d_{\eta\sigma^\prime}e^{-\tau_1 H_{QD}} d^\dag_{\beta\sigma} e^{-(\beta-\tau-\tau_2)H_{QD}}d^\dag_{\gamma\sigma^\prime} e^{-\tau_2 H_{QD}} d_{\alpha\sigma} e^{-(\tau-\tau_1) H_{QD}} \Bigr )\Bigr ] 
	\notag \\
	 - & \int\limits_0^{\tau} d\tau_1 \int\limits_0^\tau d\tau_2 \, \Bigr [ n_F(E) e^{(\tau_1-\tau_2)E} \Tr \Bigl ( 
	 d^\dag_{\gamma\sigma^\prime}e^{-\tau_2 H_{QD}} d^\dag_{\beta\sigma} e^{-(\beta-\tau)H_{QD}} d_{\alpha\sigma} e^{-(\tau-\tau_1) H_{QD}} d_{\eta\sigma^\prime} e^{-(\tau_1-\tau_2) H_{QD}} \Bigr ) 
	 \notag \\
	& + [1-n_F(E)] e^{-(\tau_1-\tau_2)E} \Tr \Bigl ( 
	d_{\eta\sigma^\prime} e^{-\tau_2 H_{QD}} d^\dag_{\beta\sigma} e^{-(\beta-\tau)H_{QD}} d_{\alpha\sigma} e^{-(\tau-\tau_1) H_{QD}} d^\dag_{\gamma\sigma^\prime} e^{-(\tau_1-\tau_2) H_{QD}} \Bigr ) \Bigr ]
	\notag \\
  - & \int\limits_0^{\beta-\tau} d\tau_1 \int\limits_0^\tau d\tau_2 \, \Bigr [ n_F(E) e^{(\tau_1-\tau_2)E} \Tr \Bigl ( 
	 d^\dag_{\gamma\sigma^\prime}e^{-\tau_2 H_{QD}} d_{\alpha\sigma} e^{-\tau H_{QD}}  d^\dag_{\beta\sigma}e^{-(\beta-\tau-\tau_1) H_{QD}} d_{\eta\sigma^\prime} e^{-(\tau_1-\tau_2) H_{QD}} \Bigr ) 
	 \notag \\
	& +  [1-n_F(E)] e^{-(\tau_1-\tau_2)E} \Tr \Bigl ( 
	 d_{\eta\sigma^\prime}e^{-\tau_2 H_{QD}} d_{\alpha\sigma} e^{-\tau H_{QD}}  d^\dag_{\beta\sigma}e^{-(\beta-\tau-\tau_1) H_{QD}} d^\dag_{\gamma\sigma^\prime}  e^{-(\tau_1-\tau_2) H_{QD}} \Bigr )\Bigr ] 
	 \Biggr \}
 .
\end{align}	
Now we perform integration over imaginary times $\tau_1$ and $\tau_2$, neglect all terms which are exponentially small in $\beta E_c$, and make analytic continuation to real frequency. Then we find	
\begin{align}
\mathcal{A}_{\rm tot}^\sigma(\varepsilon) & =   \pi \sum_{\alpha\beta\gamma\eta} \sum_{i,f,\sigma^\prime} \int d\varepsilon^\prime Q^\sigma_{\beta\alpha}(\varepsilon)Q^{\sigma^\prime}_{\gamma\eta}(\varepsilon^\prime) p_i \frac{1+e^{-\beta\varepsilon}}{1+e^{-\beta \varepsilon^\prime}} \Biggl [ 
\langle i | d^\dag_{\gamma \sigma^\prime} \frac{1}{\varepsilon-E_f + H_{QD}} d_{\alpha\sigma} +
d_{\alpha\sigma} \frac{1}{\varepsilon+E_i -  H_{QD}} d^\dag_{\gamma \sigma^\prime}| f\rangle 
 \notag \\
\times &
\langle f | d^\dag_{\beta \sigma} \frac{1}{\varepsilon-E_f + H_{QD}} d_{\eta\sigma^\prime} + 
d_{\eta\sigma^\prime} \frac{1}{\varepsilon+E_i -  H_{QD}} d^\dag_{\beta \sigma}| i\rangle 
\delta(\varepsilon^\prime +E_f-\varepsilon-E_i)
 + e^{\beta\varepsilon} \delta(\varepsilon^\prime +E_f-\varepsilon-E_i)
 \notag \\ \times & 
\langle i | d^\dag_{\beta \sigma} \frac{1}{\varepsilon-E_i + H_{QD}} d^\dag_{\gamma\sigma^\prime} + 
d^\dag_{\gamma\sigma^\prime} \frac{1}{\varepsilon+E_f -  H_{QD}} d^\dag_{\beta \sigma}| f\rangle 
\langle f | d_{\eta \sigma^\prime} \frac{1}{\varepsilon-E_i + H_{QD}} d_{\alpha\sigma} + 
d_{\alpha\sigma} \frac{1}{\varepsilon+E_f -  H_{QD}} d_{\eta \sigma^\prime}| i\rangle  \Biggr ] .
\label{app4}
\end{align}
\end{widetext}
We note that this expression can be derived in a different approach (see Appendix \ref{App2}). This gives a transparent interpretation of $i$ and $f$ states as initial and final states of a quantum dot. The inelastic scattering will correspond to different initial and final states. With this in mind, we mention that the second contribution  in Eq. \eqref{app4} contains the initial and final states which differ by the number of electrons. Such contribution does not correspond to the scattering process and we omit it. Finally, we obtain Eq. \eqref{eq:4}.


\section{Relation of $\mathcal{A}_{\rm inel}^\sigma(\varepsilon)$ to the inelastic cotunneling rate}
\label{App2}

In this appendix we demonstrate relation between the quantity $\mathcal{A}_{\rm inel}^\sigma(\varepsilon)$ and the inelastic cotunneling rate computed by means of the generalized T-matrix approach. The quantum mechanical rate for the transition from the eigenstate $|I\rangle$ to the eigenstate $|F\rangle$ of the Hamiltonian $H_{QD}+H_{R}$ due to the presence of $H_T$ is given as
\begin{equation}
\Gamma_{I\to F} = 2\pi \bigl | \langle I | \texttt{\large T} | F\rangle \bigl |^2 \delta(E_I-E_F), 
\end{equation}
where 
\begin{equation}
\texttt{\large T}  = H_T + H_T \frac{1}{E_I-H_{QD}-H_{R}}H_T + \dots
\end{equation}
This expression is the so-called generalized Fermi Golden rule in the T-matrix approach (see Ref. [\onlinecite{Timm2008}] and references therein). The 4th order contribution, which we call cotunneling rate, is
\begin{equation}
\Gamma_{I\to F}^{IC} = 2\pi \bigl | \langle I | H_T \frac{1}{E_I-H_{QD}-H_{R}}H_T | F\rangle \bigl |^2 \delta(E_I-E_F)
\end{equation}
We choose the following initial state $|I\rangle = |i\rangle |FS\rangle$ where $|FS\rangle$ denotes the Fermi sea in the reservoir. There are two-possible final states  $|F_+\rangle = |f\rangle a^\dag_{k_2\sigma_2} a_{k_1\sigma_1} |FS\rangle$ and $|F_-\rangle = |f\rangle  a_{k_1\sigma_1} a^\dag_{k_2\sigma_2}  |FS\rangle$ with additional electron-hole pair. The corresponding intermediate states are
$|V_+\rangle = |v_+\rangle a_{k_1\sigma_1} |FS\rangle$ (with additional electron on the quantum dot)   and
$|V_-\rangle = |v_-\rangle a^\dag_{k_2\sigma_2}  |FS\rangle$  (with additional hole on the quantum dot). Then we find
\begin{widetext}
\begin{gather}
\langle F | H_T \frac{1}{E_I-H_{QD}-H_{R}}H_T | I\rangle  = \sum
  \overline{t}_{p\alpha} t_{\beta p^\prime} \Biggl \{
\frac{\langle f| \langle FS| a^\dag_{k_1\sigma_1} a_{k_2\sigma_2} a^\dag_{p\sigma} d_{\alpha\sigma}  a_{k_1\sigma_1}|FS\rangle|v_+\rangle \langle v_+|\langle FS|a^\dag_{k_1\sigma_1} d^\dag_{\beta\sigma^\prime}a_{p^\prime\sigma^\prime}|FS\rangle |i\rangle}{E_i - E_{v_+}+\varepsilon_{k_1\sigma_1}} \notag \\
 +  
\frac{\langle f| \langle FS| a_{k_2\sigma_2} a^\dag_{k_1\sigma_1}  d^\dag_{\beta\sigma^\prime}  a_{p^\prime\sigma^\prime} a^\dag_{k_2\sigma_2}|FS\rangle|v_-\rangle \langle v_-|\langle FS|a_{k_2\sigma_2} a^\dag_{p\sigma}d_{\alpha\sigma}|FS\rangle |i\rangle}{E_i - E_{v_-}-\varepsilon_{k_2\sigma_2}}\Biggr \}
=\sum\limits_{\alpha\beta}\overline{t}_{k_2\alpha} t_{\beta k_1} n_{k_1\sigma_1} (1-n_{k_2\sigma_2})\notag \\
\times \Biggl \{ n_{k_1\sigma_1} 
\Bigl \langle f| d_{\alpha\sigma_2}\frac{1}{E_i - H_{QD}+\varepsilon_{k_1\sigma_1}} d^\dag_{\beta\sigma_1}|i\Bigr \rangle 
-
 (1-n_{k_2\sigma_2})
\Bigl\langle f| d^\dag_{\beta\sigma_1}\frac{1}{E_i - H_{QD}-\varepsilon_{k_2\sigma_2}}d_{\alpha\sigma_2}|i\Bigr \rangle \Biggr \}  .
\end{gather}
Here $n_{k\sigma}$ is the particle number operator for the state in the reservoir.
Hence, performing the thermal average over reservoir states, we obtain  the rate from the state $|i\rangle$ to the state $|f\rangle$ of the quantum dot hamiltonian $H_{QD}$ as 
\begin{gather}
\Gamma_{i\to f}^{IC}  = 2\pi \int d\varepsilon d\varepsilon^\prime n_F(\varepsilon)(1-n_F(\varepsilon^\prime)) 
\sum_{\alpha\beta\gamma\eta\sigma_1\sigma_2} Q^{\sigma_2}_{\beta\eta}(\varepsilon)Q^{\sigma_1}_{\gamma\alpha}(\varepsilon^\prime)\Bigl \langle i\Bigl | 
d^\dag_{\gamma\sigma_2}\frac{1}{\varepsilon^\prime-E_i +H_{QD}}d_{\eta\sigma_1}
+d_{\eta\sigma_1}\frac{1}{E_i +\varepsilon - H_{QD}} 
 d^\dag_{\gamma\sigma_2}\Bigr |f\Bigr \rangle 
\notag \\
\times 
\Bigl \langle f\Bigl | 
d^\dag_{\beta\sigma_1}\frac{1}{\varepsilon^\prime-E_i +H_{QD}}d_{\alpha\sigma_2}
+d_{\alpha\sigma_2}\frac{1}{E_i +\varepsilon - H_{QD}} 
 d^\dag_{\beta\sigma_1}\Bigr |i\Bigr \rangle 
 \delta(\varepsilon+E_i-E_f-\varepsilon^\prime) .
\end{gather}
\end{widetext}
Performing thermal averaging over the initial states of the quantum dot, we obtain
\begin{equation}
\Gamma^{IC} = \sum_{i\neq f} p_i \Gamma_{i\to f}^{IC} =  \int d\varepsilon  \frac{\sum\limits_\sigma \mathcal{A}_{\rm inel}^\sigma(\varepsilon)}{2\cosh^2(\beta\varepsilon/2)}
\end{equation}

\bibliography{biblio}

\begin{thebibliography}{53}
\bibitem [{foo()}]{footnote1}%
  \BibitemOpen
  \href@noop {} {}\bibinfo {note} {We assume that $Q_{\alpha\beta} \propto
  \phi_{\alpha}\bar{\phi}_\beta$ where $\phi_\alpha$ is a random vector with
  Gaussian statistics obeying the following relations: $\langle \phi_\alpha
  \bar{\phi}_\beta\rangle = Q \delta_{\alpha\beta}$ and $\langle \phi_\alpha
  {\phi}_\beta\rangle = (2/\bm{\beta}-1) Q \delta_{\alpha\beta}$ (see Ref.
  [\onlinecite{ABG}]).}\BibitemShut {Stop}%
\end{thebibliography}

\end{document}